\documentclass{jstatphys}
\usepackage[reqno,sumlimits]{amsmath}
\usepackage{amsfonts}
\usepackage{amssymb}
\usepackage{bbm}
\usepackage{epsfig}

\newcommand{\bc}{\begin{center}}
\newcommand{\ec}{\end{center}}

\newcommand{\be}{\begin{equation}}
\newcommand{\ee}{\end{equation}}
\newcommand{\ba}{\begin{eqnarray}}
\newcommand{\ea}{\end{eqnarray}}

\newcommand{\bi}{\begin{itemize}}
\newcommand{\ei}{\end{itemize}}

\begin{document}


\title[2D Crystal Shapes, Droplet Condensation and Exponential Slowing Down]
{2D Crystal Shapes, Droplet Condensation and Exponential Slowing Down in 
Simulations of First-Order Phase Transitions}

\author{Thomas~Neuhaus%
\footnote{Finkenweg 15, D -- 33824 Werther, Germany;
e-mail: neuhaus@rock.helsinki.fi.}
and Johannes~S.~Hager%
\footnote{Institute for Physical Science and Technology,
University of Maryland, College Park, Maryland 20742.}$^,$
\footnote{Present address: Fachbereich Physik, Universit\"at Essen, D-45117 Essen,
Germany; email: johannes@theo-phys.uni-essen.de}
}

\runningauthor{Neuhaus and Hager}

\date{Version of \today}

\begin{abstract}

Multicanonical ensemble simulations for the simulation of first-order phase
transitions 
suffer from exponential slowing down. Monte Carlo autocorrelation 
times diverge exponentially with free energy barriers 
$\Delta F$, which in $L^d$ boxes grow as $L^{d-1}$. We exemplify the 
situation in a study of the 2D Ising-model at temperature  
$T/T_c=0.63$ for two different lattice manifolds, toroidal lattices 
and surfaces of cubes. For both geometries the effect 
is caused by discontinuous droplet shape transitions between
various classical crystal shapes obeying geometrical constraints. 
We use classical droplet theory and numerical simulations
to calculate transition points and barrier heights. 
On toroidal lattices we determine finite size corrections to
the droplet free energy, which are given by a linear combination of 
Gibbs-Thomson
corrections, capillary wave fluctuation corrections, constant terms 
and logarithmic terms in the droplet volume. Tolman corrections are absent. 
In addition, we study the finite size effects on the condensation phase 
transition, which occurs in infinite systems at the 
Onsager value of the magnetization. We find that this transition is of 
discontinuous order 
also. A combination of classical droplet theory and Gibbs-Thomson corrections 
yields a fair description for the transition point and for 
the droplet size discontinuity for large droplets. We also estimate the 
nucleation barrier that has to be surmounted in the formation of the stable 
droplet at coexistence.

\end{abstract}

\keywords{Nucleation, Ising model, Multicanonical simulations, geometric  
phase transitions.}

\section{Introduction}

First-order phase transitions play an important role 
in many branches of physics ranging from the well known 
liquid vapor transition \cite{Bin80,Bin87}
to nuclear physics \cite{Ric00}, protein folding 
\cite{dgo96} or even to the symmetry breaking in the 
early universe \cite{a99}. Even for the intensely 
studied liquid vapor transition there are still 
considerable uncertainties in the calculations of the
decay rate of metastable states \cite{Bin80}, partially
due to  unknown finite size and finite curvature 
corrections to the free energy and the surface tension 
of droplets. 

The present study uses the two dimensional
Ising-model to investigate the influence of such finite size 
effects, because it is conceptually simple and there 
is a large body of rigorous results, which can
be used in the comparison to simulation data. The 
partition function of the Ising-model is given by
\be
Z = \sum_{conf.}e^{ -\beta H} \quad,
\end{equation}
with the Hamiltonian
\be
H = H_I - hM 
:= - \sum_{<i,j>}{s_is_j} -h\sum_{i}s_i \quad,
\end{equation}
where $H_I$ contains the usual nearest neighbor interaction and the 
magnetization $M$ couples linearly to a external magnetic field $h$.
We use the multicanonical sampling method to study the model in the 
whole magnetization interval $[-L^2,L^2]$ at the inverse temperature
$\beta=0.7$(corresponding to $T/T_c=0.63$, with $\beta_c=\ln(1+\sqrt{2})/2$), 
which is sufficiently low to pronounce effects due to 
first-order phase transitions, but still high enough to use a isotropic 
surface free energy as a good first approximation.
Most of our efforts will be focused on the magnetic probability distribution
\be
P_L(M) = \frac{1}{Z}\sum_{conf.}e^{ -\beta H}\delta(M-\sum_{i}s_i) \quad,
\end{equation}
which up to a normalization factor equals the restricted partition function  
$Z(m,L)$.
The distribution $P_L(M)$ was already studied in \cite{s80,b81,fb82,lnr95}
with the aim to understand
the dynamics of the decay of a metastable state. Our goal in this paper is to 
produce a quantitative description for high resolution Monte-Carlo data of $P_L(M)$
in terms of classical droplet theory, including the leading finite-size effects. 

The paper is organized as follows. In section 2 we give a brief review of 
the multicanonical sampling method and present numerical evidence of residual 
exponential slowing down, due to the singularities associated with first 
order phase transitions. We argue that similar limitations are present 
in other broad histogram sampling methods. In section 3 we present the 
classical theory for the boundary induced geometric phase transitions in 
toroidal and cube-surface geometries. Section 4 discusses the influence 
of a variety of possible finite size effects. Section 5 gives a detailed 
account of our simulation data for both geometric phase transitions and the 
droplet condensation transition, including the analysis of finite size effects.
In section 6 we conclude our findings.

\section{Multicanonical Method}\label{mucamethod}

Multicanonical (Muca) sampling was invented \cite{bn91,bn92} 
to eliminate the exponential slowing down of canonical 
(Metropolis or heat bath) simulations near temperature-
or field-driven first-order phase transitions. 
At inverse temperatures $\beta>\beta_c$ the magnetic 
probability distribution $P_L(M)$ as a function of the 
magnetization $M$ in the Ising-model has two 
maxima, which we denote by $\pm M_L^{max}$,
separated by a valley, where the probability is 
suppressed by a factor $e^{-\Delta F}$, 
due to the additional free energy 
$\Delta F\sim\sigma L$ of the interface present in 
the two phase region. Already for moderate system 
sizes canonical simulations are not able to sample
these exponentially suppressed states and the 
simulation gets trapped in one of the maxima of 
$P_L(M)$. Similar problems arise in simulations of 
spin glasses. The Muca method remedies this issue by 
biasing the sampling with a weight factor 
$P_L(M)^{-1}$, thereby producing flat histograms
\cite{bhn93a,bhn93b}. Alternatively one can use the 
inverse of the density of states $n(E)^{-1}$
for a given internal energy $E$ to enhance the 
sampling of the suppressed states in energy driven first
order phase transitions.  Thus the simulations are performed
with an effective Hamiltonian
$H_{eff}=H_I+\beta^{-1}\ln P_L(M)$.
Unbiased averages
can be calculated  via
\be
<\!A\!>\, = \frac{\sum_{conf.}A\, e^{ -\beta (H_I-H_{eff})}}{\sum_{conf.}
e^{ -\beta (H_I-H_{eff})}} \quad.
\end{equation}
A major practical problem in the application of the 
method is, that one needs a fairly good approximation
of $P_L(M)$ to run a efficient Muca simulation. 
The determination of an estimate for $P_L(M)$ by 
conventional importance sampling may consume already
large parts of the overall computation time, 
especially for systems with a rough energy 
landscape, like spin-glasses. Recursive 
schemes have been proposed \cite{b96} for the cases 
where finite size scaling cannot be used to extrapolate
$P_L(M)$ to large system size $L$.
Recently several new ideas which tackle
this practical problem by forcing the simulation 
into the unfavorable states have been put 
forward \cite{lw01,n01}.

We now want to focus on the performance of the 
algorithm, supposing we have the exact probability 
distribution $P_L(M)$ at hand. For this purpose we
study the system size dependence of the 
autocorrelation time $\tau$, defined as the
mean time needed by the algorithm to go from the 
left side of the magnetization range at $-M_L^{max}$ 
to the right one at $M_L^{max}$ and back
again. After transforming away the energy barrier due 
to the interface we may at best expect a random walk 
motion of the algorithm in the magnetization. A walk 
between $-M$ and $M$ then takes $O(M^2)$ spin flips. 
With $M\sim L^2$ and the usual definition of 
Monte-Carlo time in units of $L^2$ attempted spin
flips (sweeps) we arrive at $\tau\sim L^2$ for the optimal 
behavior of $\tau(L)$.
\begin{figure}[tb]
\centerline{ \begin{minipage}[c]{13.2cm}
(a) \psfig{file=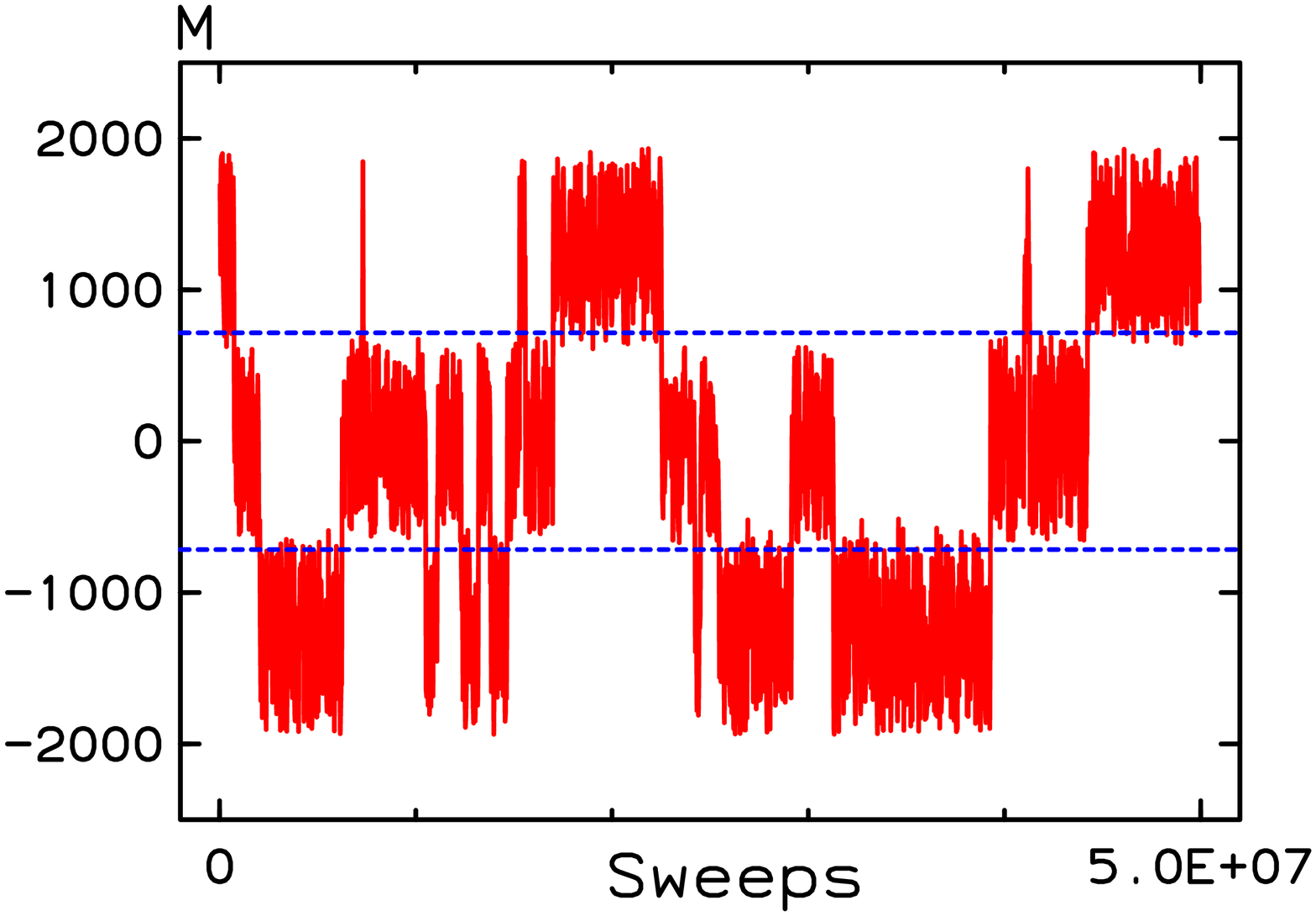 ,angle=270,width=5.5cm}
(b) \psfig{file=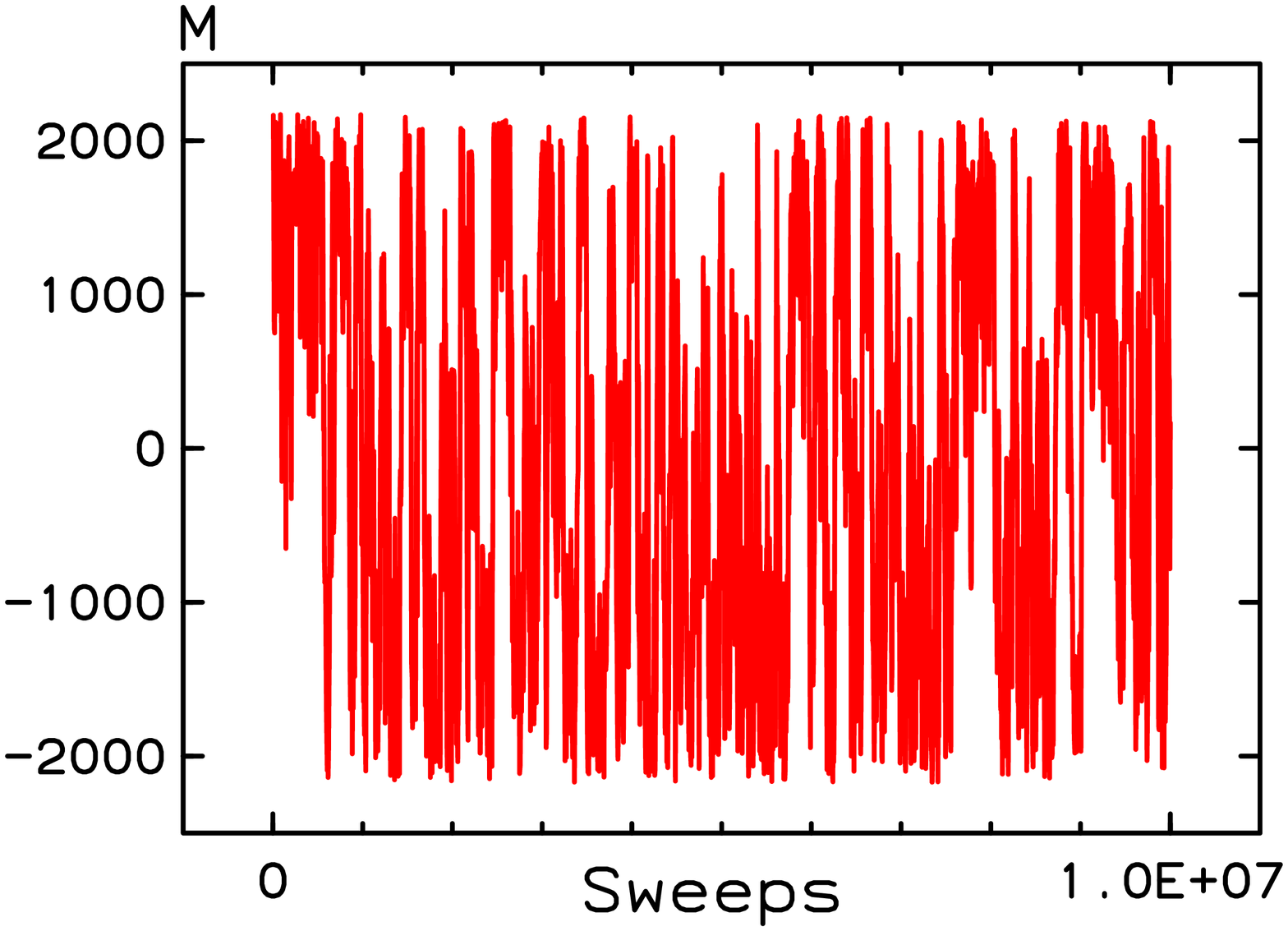,angle=270,width=5.5cm}
\end{minipage}}
\caption[a]{(a) Time series for magnetizations 
$M$ in Muca simulations at $\beta=0.7$ on a toroidal $44^2$ lattice 
as compared to (b) time series for $M$ on a cube-surface 
SH(20) lattice, which has a similar volume.
The statistics is $5\cdot10^7$ sweeps on the torus and $10^7$ 
sweeps on the SH lattice. The horizontal lines for the 
torus indicate the position of shape transitions.
\label{fig:time_series} } \end{figure}
The sobering news of Fig.~\ref{fig:time_series}
-- displaying time series for the magnetization $M$ 
in Muca simulations of the 2D Ising-model at 
inverse temperature $\beta=0.7$ on a torus and a
cube-surface (see Fig.~\ref{surface_hypercube}) -- is, that the 
assumption of random walk like motion in the whole 
magnetization interval is plainly wrong. For the torus 
geometry (pbc.) one clearly sees that the 
interval for the magnetization at least is divided into three 
sectors with random walk like behavior, separated by 
two barriers. For the cube-surface, which we abbreviate in the
sequel as SH (for surface of a hypercube in case $d>2$), the
presence of barriers is less obvious. 
An estimate of the magnitude of barriers
can be obtained, if $\tau$ is fitted with the form
\begin{equation}
\tau= A_\tau V e^{R \sigma \partial\Omega_{max} } ,
\end{equation}
where the maximal surface $\partial\Omega_{max}$, which
determines the exponential slowing down, is $2L$ on a 
toroidal $L^2$ box and $4(L-1)$ on a $SH(L)$ lattice. 
The symbol $\sigma$ denotes the surface tension and $V$ is the volume. 
We obtain the values
\begin{eqnarray}
R &=&  0.121(14)\quad \mbox{Torus} \label{bsds} \\
R &=&  0.031(05)\quad \mbox{SH} \label{eq:r_value} \quad,
\end{eqnarray}
from the fits as displayed in Fig.~\ref{fig:tautimes}. 
Into the fit enter data, which in the figure
are displayed with solid symbols. The measured $R$-values are
significantly smaller than values $R \approx 1$ as expected for non 
multicanonical simulations, but clearly indicate the presence of residual 
exponential slowing down.
\begin{figure}[t]
\begin{center}
\epsfig{file=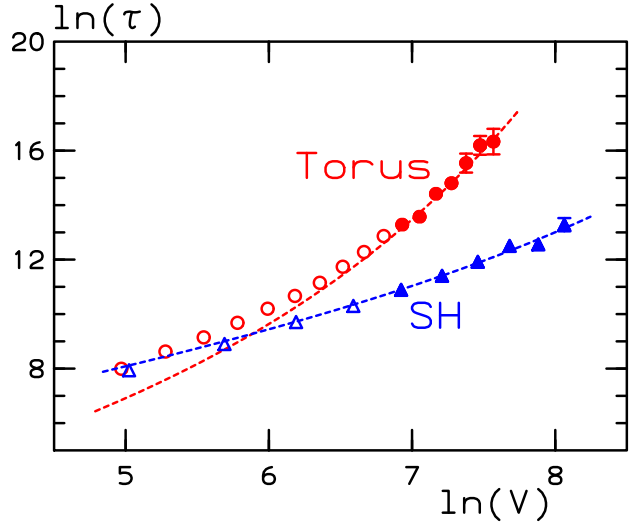,width=08cm,angle=270}
\caption{
Multicanonical autocorrelation times $\tau$ for the torus (circles)
and SH lattices (triangles) in the 2D Ising-model at $\beta=0.7$
in units of heatbath sweeps as a function of the box 
volume $V$. The plot is double logarithmic. Curves are 
explained in the text. } \label{fig:tautimes} \end{center}
\end{figure}
We show, that the barrier on the torus, which leads 
to a exponential slowing down, is caused by a 
geometrically induced first-order transition from a 
droplet to a strip domain, whose barrier value can
be calculated using classical droplet theory. The
droplet theory result $R=0.1346...$ agrees within error bars with
the measured value of (\ref{bsds}). A superficial 
inspection of the time series for the cube-surface, 
see Fig.~\ref{fig:time_series}b, might lead to the 
guess, that in this case no barriers are present. We will show, that 
in this case there is a series of three different discontinuous 
transitions in-between phase space regions, where droplets occupy one, 
two, three and four corners on the cube-surface. The 
barrier values are $R_{1/2}=0.02987$, $R_{2/3}=0.02977$ and 
$R_{3/4}=0.03441$, which all are smaller than corresponding barriers on 
the torus and, the maximum value $R_{3/4}$ again agrees within error bars 
with the measured value cited in (\ref{eq:r_value}).

Now, as already pointed out by Leung and Zia 
\cite{lz90}, one could avoid this type of geometrical 
phase transitions by simulating the Ising-model on the
surface of a sphere, and actually the cube-surface 
may be viewed as a crude approximation to 
the sphere. Alas there is no way known to put a 
regular lattice of arbitrary volume on the surface 
of a sphere, but off lattice simulation of liquid-gas
systems could do the job. Are we then able to achieve 
random walk like Muca dynamics in the whole interval of 
magnetizations ? Our answer to this 
question still is no, since there is an additional
first-order transition present, namely the droplet 
condensation phase transition from a uniform one phase 
region to the phase separated two phase region, where 
we have yet another essential singularity in the free energy. 
Here again we find a, albeit much smaller barrier for 
the Muca simulations, which is now directly related to
the physical nucleation barrier associated with the 
formation of a critical nucleus \cite{f67,l67}. Now, 
what at first sight just looks like an embarrassing
limitation of the algorithm, at a second thought gives 
us interesting information about the droplet 
condensation phase transition. In the quantitative
analysis of this data Gibbs-Thomson corrections and other finite size 
effects will play a prominent role, since 
the volume of the coexisting droplet scales 
for large system sizes asymptotically as $L^{4/3}$  
and thereby introduces an additional scale.
\begin{figure}[t]
\begin{center}
\epsfig{file=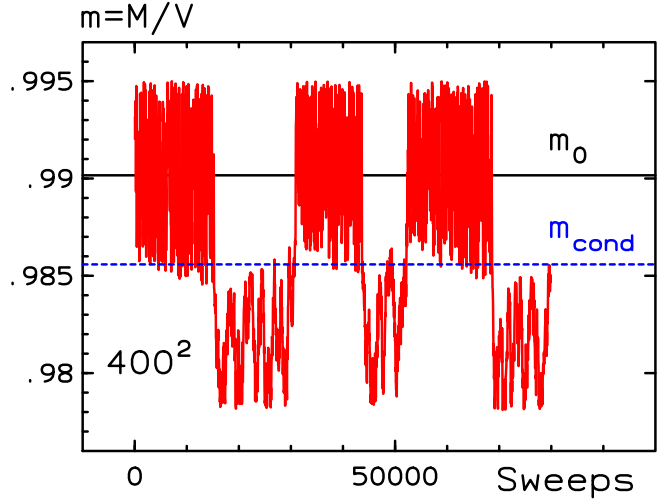,width=08cm,angle=270}
\caption{
Time series for magnetization densities 
$m$ in Wang-Landau density of states simulations at 
$\beta=0.7$ on a toroidal $400^2$ lattice and at
$f=1.00007$. The simulation slows down at the position
of the droplet condensation phase transition $m_{cond}$, which
we study in the present paper.
        }
\label{fig:landau_wang}
\end{center}
\end{figure}
Our findings not only are relevant for the algorithmic 
performance of Muca simulations, but most likely limit the
performance of other broad histogram methods -- like the recently
introduced Wang and Landau density of 
states sampling method\cite{lw01} -- as well, provided
the Monte-Carlo covers a portion of phase space, which contains one of
the mentioned singularities. For purposes of illustration 
we display in Fig.~\ref{fig:landau_wang} the magnetization density time 
series obtained from a Wang-Landau broad histogram sampling
in the magnetization of the 2D Ising-model on a $400^2$
lattice at $\beta=0.7$.  The value of the Wang-Landau parameter
$f$ is $f=1.00007$ and the magnetization density of states is updated 
with a heat bath algorithm. 
The horizontal lines in the figure denote the Onsager value $m_0$ 
and the position $m_{cond}$ of the droplet condensation phase 
transition. The existence of a barrier is indicated by the presence of 
flip flops between two different regions of the 
phase space, below and above $m_{cond}$. This exponential slowing down
worsens, if either lattice sizes are increased, or $f$ is tuned
to the value unity.

\section{Classical droplets}\label{classic}

Below the critical point the bulk density of the 
spontaneous magnetization $m_0(T)$ in a infinite system 
is given by Onsagers solution \cite{Mcc73}, which at $\beta=0.7$
predicts the value $m_0=0.99016...$ for the magnetization density. If we 
restrict the mean density of the magnetization to some value in the 
interval $[-m_0,m_0]$ the system separates 
into two phases with magnetizations $\pm m_0(T)$ divided 
by an interface. The orientation dependent free energy 
of a interface can be obtained from the spin-spin correlation 
function via a dual transformation \cite{Rot84}. The equilibrium shape 
of a droplet with volume $\Omega$ of the minority phase, embedded in the 
majority phase, can be obtained by the celebrated Wulff construction 
\cite{w01}. The total free energy of the droplet is given by \cite{lz90}
\be \label{exact1}
\Sigma_D = 2\sqrt{W\Omega}
\end{equation}
where
\be \label{wulffshape}
W = \frac{4}{\beta^2}\int_{0}^{\beta\sigma_0}dx \cosh^{-1}\left[
\frac{\cosh^2(2\beta J)}{\sinh(2\beta J)}-\cosh(x)\right]
\end{equation}
is the volume bounded by the Wulff plot of the 
orientation dependent surface free energy, and 
$\sigma_0(T)=2J+\frac{1}{\beta}\ln\tanh\beta J$ is the 
free energy of the $(1,0)$ surface, which at $\beta=0.7$
has the value $\sigma_0=0.89643...$~. In the presence of 
boundaries different equilibrium shapes can occur 
\cite{w67,zat88}. In the case of a torus with periodic
boundary conditions there is a first-order phase transition from the droplet 
shape - as given by the Wulff construction - to a strip with surface
free energy $\Sigma_S=2\sigma_{0} L$. The strip is wrapped around the 
torus and the transition point is determined by the condition 
$\Sigma_D=\Sigma_S$. This was shown rigorously by Shlosman \cite{s89}. 
Employing numerical integration for the integral in 
Eq.~(\ref{wulffshape}) one can precisely calculate the transition 
point $m_{D/S}(T)$, which at $\beta=0.7$ has the value
\begin{equation}
m_{D/S} =0.36974... \quad .
\label{strip_droplet_point}
\end{equation}
In our Monte-Carlo data the transition is rounded and shifted due 
to several finite size effects, which will be discussed in detail in 
the next section.
\begin{figure}[tb]
\centerline{ \begin{minipage}[c]{13.2cm}
(a) \psfig{file=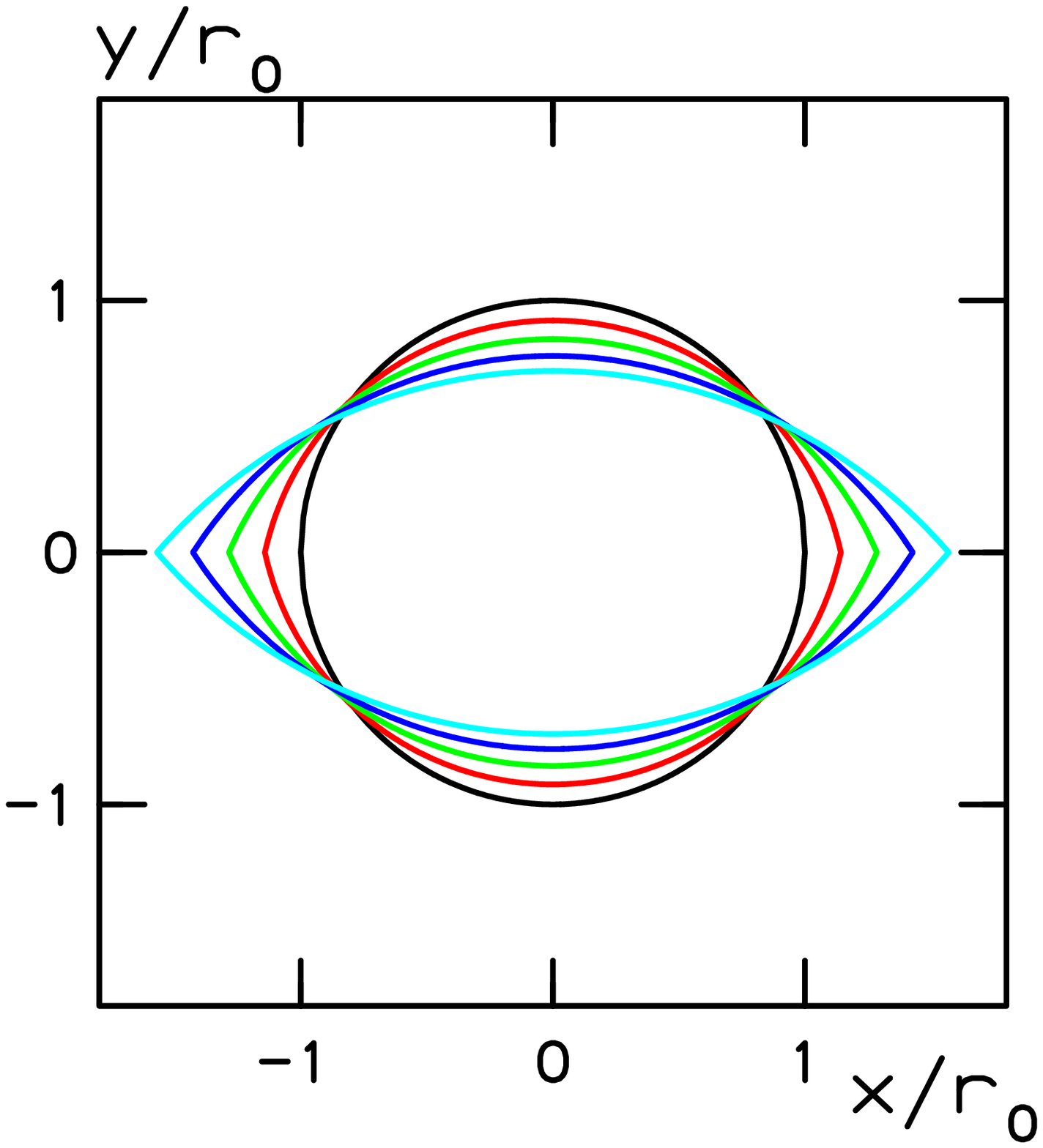,angle=000,width=5.5cm}
(b) \psfig{file=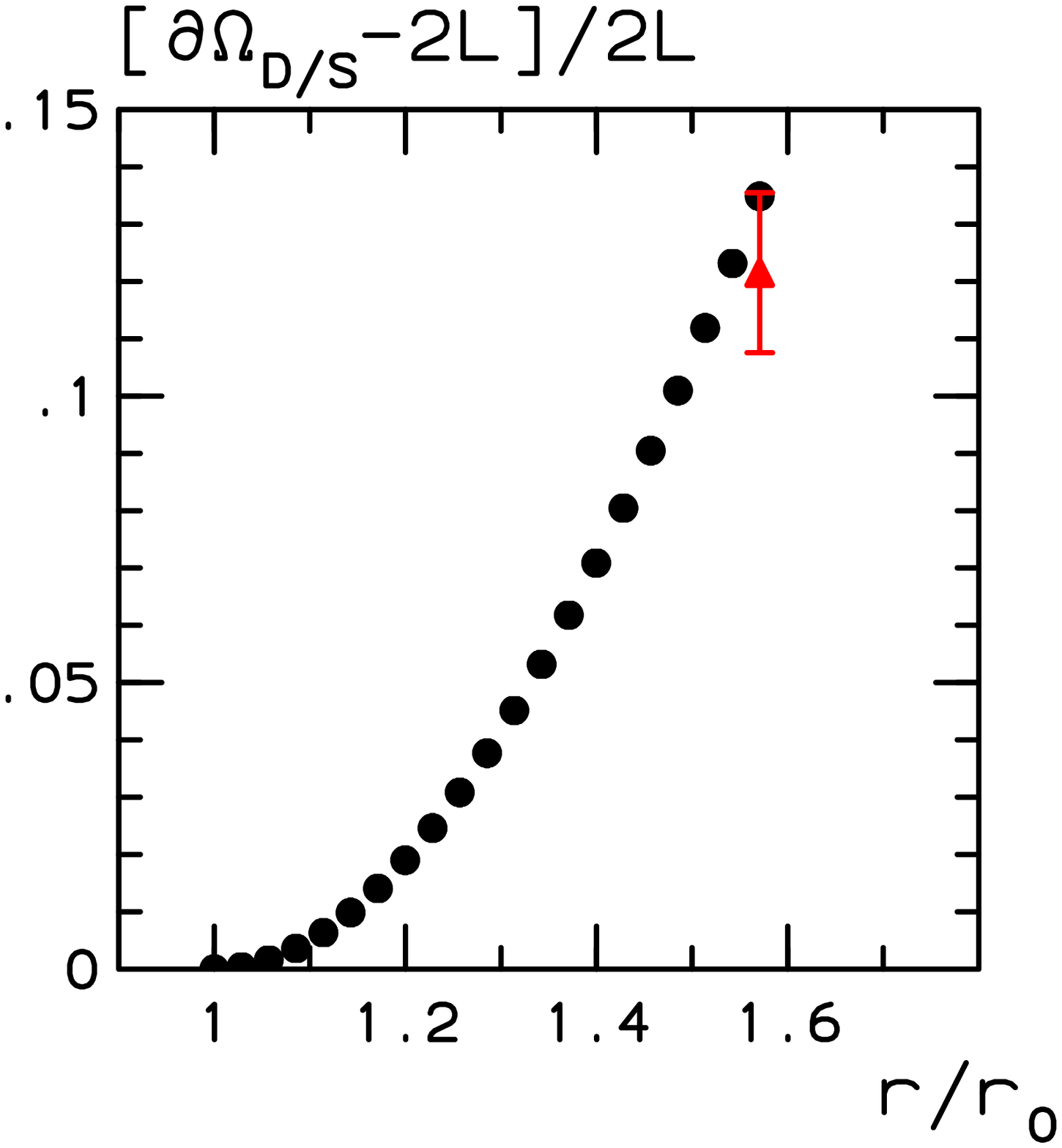    ,angle=000,width=5.5cm}
    \end{minipage}}
\caption[a]{
(a) Crystal shapes at the droplet strip transition. With the droplet volume 
fixed at its value
${L^2 / \pi}$ at coexistence 
the circular stable droplet is deformed into the unstable
saddle point lens shaped droplet. (b) The percentile increase of 
the excess length $(\partial \Omega_{D/S} -2L) /2L$ as a function
of of the half base length $r/r_0$ in units of $r_0$. At $r/r_0=\pi/2$
one reaches the saddle point configuration. The measured data point 
of (\ref{eq:r_value}) is plotted with error bars and agrees 
with the classical droplet result.}\label{fig:shapes}
\end{figure}
For the toroidal and cube-surface geometry we calculate in the rest 
of this section classical equilibrium shapes and saddle point 
configurations, which determine transition points and barrier heights 
in-between the different phases. For sake of simplicity 
we use a isotropic surface free energy, which is still a 
reasonable approximation at inverse temperature $\beta=0.7$ 
($T/T_c=0.63$), where the Monte-Carlo simulations are performed. 

Toroidal boxes of linear extent $L$ with periodic boundaries 
have a volume $V=L^2$ and without boundaries the equilibrium droplet shape is 
just a circle of radius $r=\sqrt{\Omega/\pi}$, since it minimizes the surface 
$\partial\Omega=\sqrt{4\pi\Omega}$ for a given volume $\Omega$. 
Leung and Zia \cite{lz90} argued, that the saddle 
point configuration at the transition is a lens shaped 
droplet - see Fig.~\ref{fig:shapes}a - formed by two arcs with a 
base length $L$. At the transition point $m_{D/S}/m_0=1-{2 / \pi}$
this droplet interpolates in-between a spherical droplet of radius
$r_0={L / \pi}$ and the strip. The arc shape again follows from the 
minimization of the surface at given droplet volume
${L^2 / \pi}$. Figure \ref{fig:shapes}b displays the excess length of 
the droplet, which is deformed from its spherical shape at $r/r_0=1$  
to the lens shaped saddle point configuration with half 
base length  value $r/r_0 = \pi / 2$. The excess surface of the saddle 
point configuration is \cite{lz90}
\be \label{bsd}
\frac{\partial\Omega_{D/S}-2L}{2L}
= 0.1346... \quad 
\end{equation}
and we can directly check this value in our Muca simulations, since this 
barrier gives the leading exponential contribution to the 
autocorrelation time and should equal $R$ of (\ref{bsds}). The value of 
the classical excess length is in fair agreement 
with the simulation value. Some deviations may come from the 
influence of the anisotropy of the surface free energy, which at $\beta=0.7$ 
shifts the location of the transition by a few percents.
\begin{figure}[t]
\begin{center}
\epsfig{file=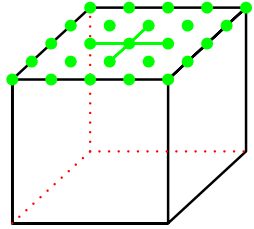,width=08cm,angle=270}
\caption{
The $SH(5)$ lattice manifold is identified with the surface
of a $5^3$ cubic box with linear extent $4$.} 
\label{surface_hypercube} \end{center}
\end{figure}
Cube-surface lattices of parameter value $L$ are 2D lattice 
manifolds denoted $SH(L)$, which are identified with the surface of a 3D
cube with linear extend $L-1$. An example is 
displayed in Fig.~\ref{surface_hypercube}, where a $SH(5)$ lattice
with linear extent $4$ is represented schematically. A $SH(L)$ lattice has a 
volume $V=6(L-2)^2+12(L-2)+8$ and there are eight sites on the corners of 
the cube, which possess three nearest neighbors instead of four.
A classical droplet configuration at magnetization $M=0$
covers four of the eight corners and has a minimal surface 
of length $\partial\Omega=4(L-1)$ separating spin up and 
spin down domains. A small droplet can lower its surface by occupying a 
corner of the cube. With increasing volume it becomes favorable to 
cover two, three and four corners and similar as on the torus there exist 
shape transitions in-between states, which occupy one, two, three and four 
corners. Due to our choice of a isotropic surface free energy all equilibrium 
droplet shapes consist out of circle segments, which are closed around the 
corners. To systematize our considerations we denote the uniquely defined 
circle segment of volume $\Omega$ and base length  $b$ by $S(b,\Omega)$ and 
its arc length by $\partial \Omega(b,\Omega)$.
\begin{figure}[tb]
\centerline{ \begin{minipage}[c]{13.2cm}
(a) \psfig{file=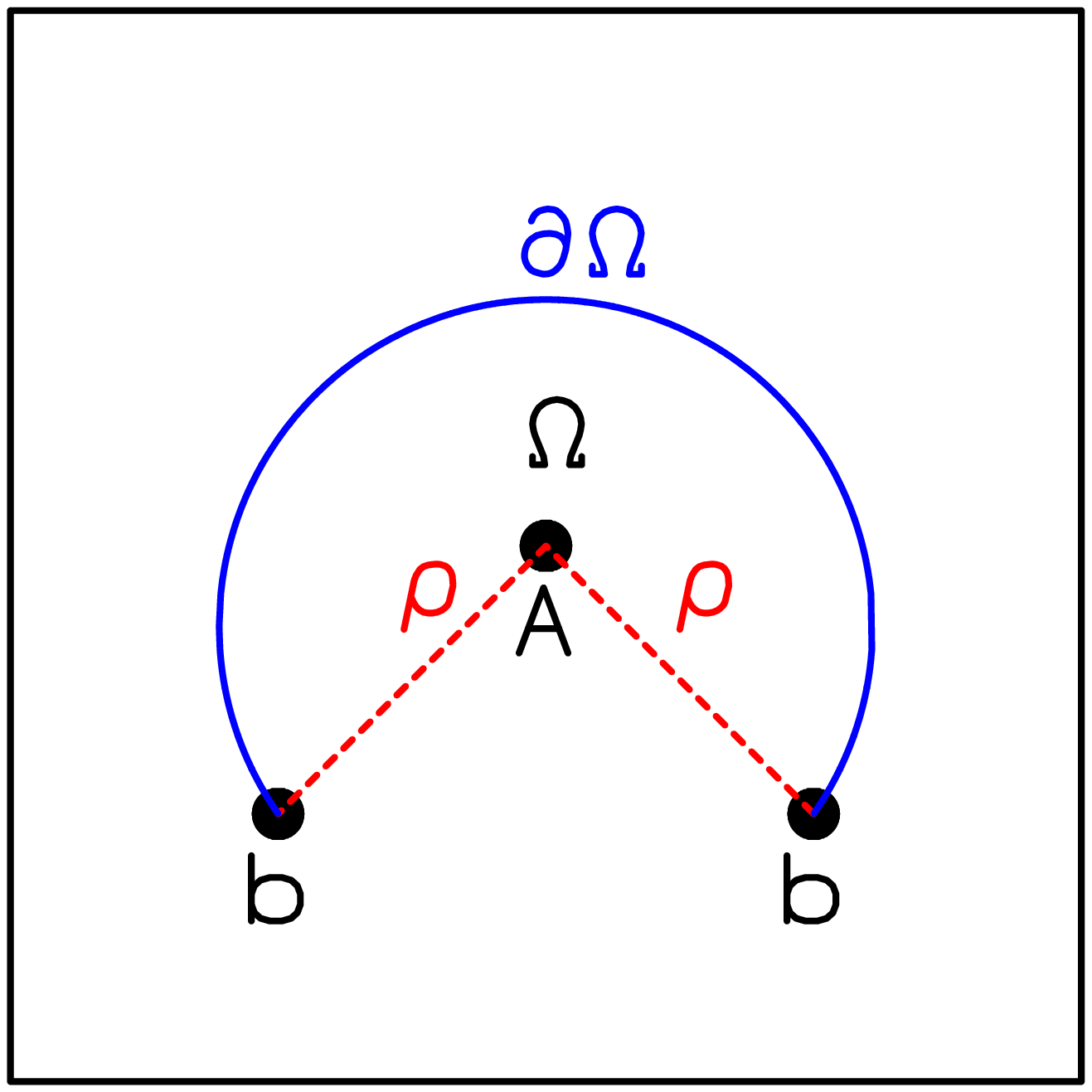,angle=000,width=5.5cm}
(b) \psfig{file=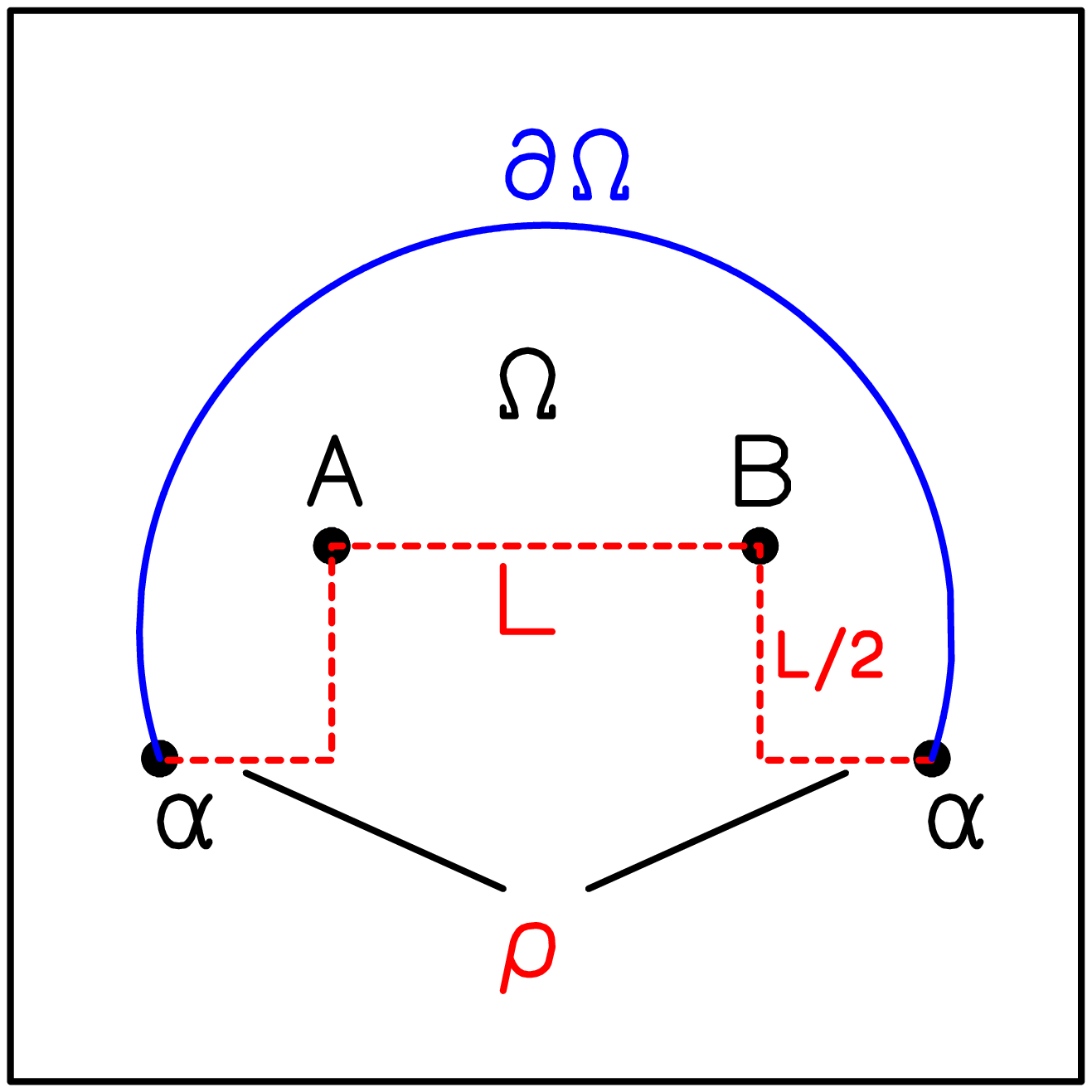,angle=000,width=5.5cm}
\end{minipage}}
\caption[a]{Generic droplet shapes on SH lattice for
(a) one corner and (b) two corner droplets. Capital letters 
denote sites on the corners of the SH lattice, small letters 
denote sites on the edges of the cube and greek letters denote sites 
neither located on edges or corners. Solid lines of the shapes 
contribute to the surface $\partial \Omega$, broken lines do not. 
Sites labeled with the same letter are to be identified.}
\label{fig:generica}  \end{figure}
\begin{figure}[t]
\begin{center}
\epsfig{file=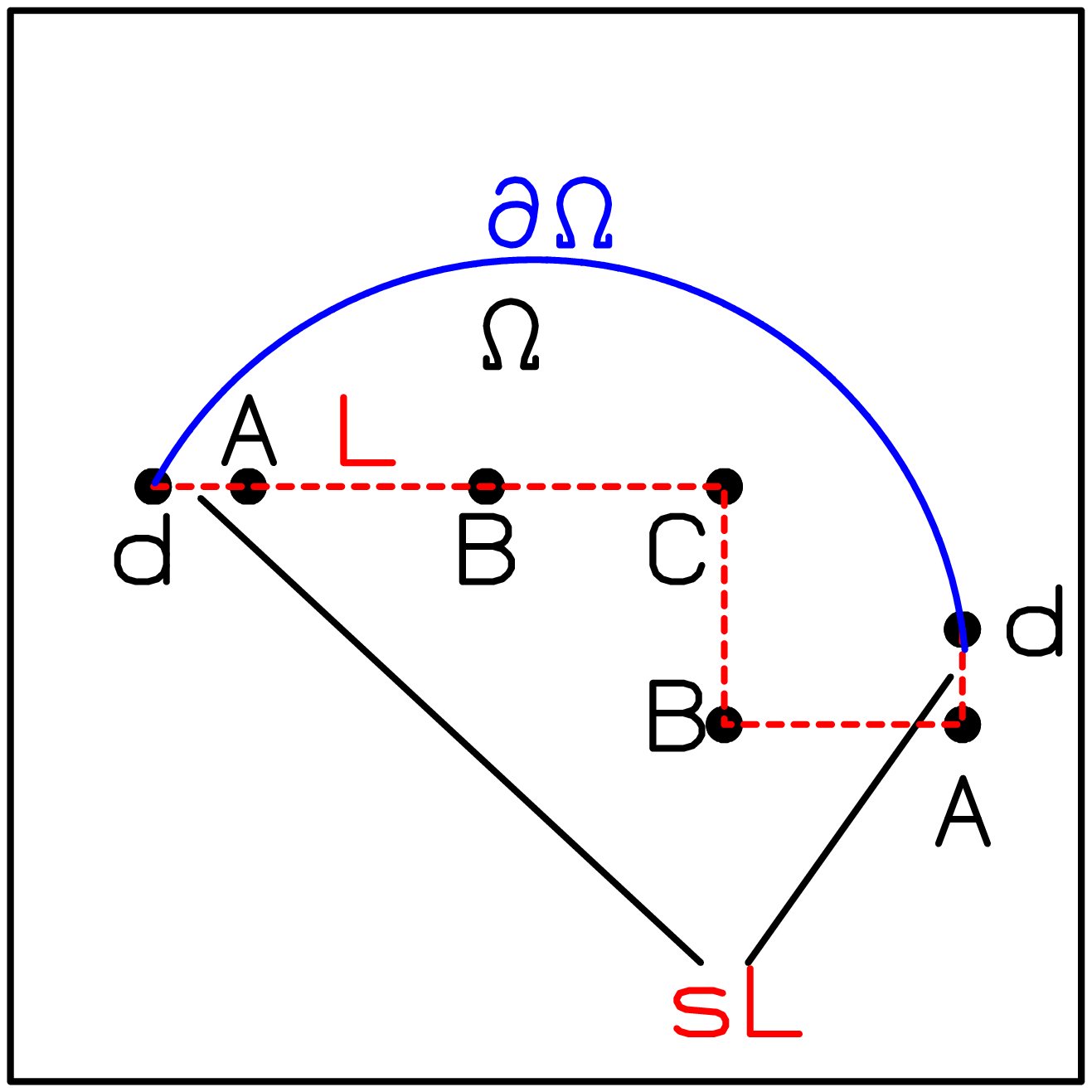,width=08cm,angle=000}
\caption{Generic droplet shape on SH lattice for
a three corner droplet.}
\label{fig:genericb}
\end{center}
\end{figure}
A droplet located at a corner has the volume 
$\Omega_1=\Omega[S(b,\Omega)]-b^2/4$ 
and the surface $\partial \Omega_1=\partial \Omega(b,\Omega)$. Note that 
the base length  itself does not contribute to the surface 
$\partial \Omega(b,\Omega)$ of the droplet.
The situation is depicted schematically in
Fig.~\ref{fig:generica}a, where the droplet covers the corner $A$ and the 
base length  has the value $b=\sqrt{2}\rho$ with $\rho$ as denoted in the 
figure. To get the equilibrium shape at constant volume $\Omega_1$ we 
minimize $\partial \Omega_1$ with respect to $b$. This results in
\be \label{surf1}
\partial \Omega_1 =\sqrt{3\pi\Omega_1} \quad
\end{equation}
as compared to $\partial \Omega =\sqrt{4\pi\Omega}$ for a face centered droplet. 
Similar considerations apply to two corner droplets and
Fig.~\ref{fig:generica}b 
displays a droplet, which covers two corners $A$ and $B$. 
The base length  $b$ has the value
$b=L+2\rho$, again with $\rho$ as denoted in the figure. For the 
non-extremal droplet we have $\Omega_2=\Omega[S(b,\Omega)]-L^2/2$
and $\partial \Omega_2 =\partial \Omega(b,\Omega)$, which again after
minimizing $\partial \Omega_2$ with respect to $b$ leads to the surface length
\be \label{surf2}
\partial \Omega_2 =\sqrt{\pi(2\Omega_2+L^2)}
\end{equation}
for the two corner equilibrium droplet of volume $\Omega_2$. Equating the 
surfaces (\ref{surf1}) and (\ref{surf2}) we find $\Omega_{1/2}=L^2$ for 
the droplet volume $\Omega_{1/2}$, where the transition from the one corner 
to the two corner droplet occurs. To calculate the barrier height at the 
transition we argue, that the saddle point configuration between the one 
corner and the two corner droplet is reached, when the site $b$ of 
Fig.~\ref{fig:generica}a occupies an additional corner: the site $B$ 
of Fig.~\ref{fig:generica}b. The volume of this droplet is $\Omega_{1/2}$ 
and its surface has the value 
$\partial \Omega_{1/2}=\partial \Omega(b=\sqrt{2}L,3/2L^2)$, 
leading to a excess length
\be
\frac{\partial\Omega_{1/2}-\partial\Omega_1}{4L}
= 0.02987 \quad
\end{equation}
normalized with the length $4L$ of the ``strip'' or four corner droplet. 
Besides this transition we find two more, from the two corner to the 
three corner droplet and from the three to the four corner droplet. 
To parameterize the possible 
shapes of a three corner droplet we introduce the parameter 
$s={\overline{Ad}}/{L}$ as indicated in Fig.~\ref{fig:genericb}. 
We read of the 
volume of the three corner droplet
\be \label{drop3}
\Omega_3=\Omega[S(b=L\sqrt{(3+s)^2+(1-s)^2},\Omega)]+L^2(1-\frac{1}{2}
(1-s)(3+s))
\end{equation}
and the surface of the droplet is again the arc length
$\partial \Omega_3(b(s),\Omega)$. Numerical minimization of 
$\partial \Omega_3$ with respect to $s$ at fixed $\Omega_3$ then gives 
us the equilibrium shape with three corners inside. Equating 
$\partial \Omega_2=\partial \Omega_3$ and 
$\partial \Omega_3=\partial \Omega_4=4L$ 
we find the volumes $\Omega_{2/3}$ and $\Omega_{3/4}$, 
where the transitions to a 
three corner droplet occur. The actual transition values $m_{1/2},m_{2/3}$ and 
$m_{3/4}$ are
\be
\frac{m_{1/2}}{ m_0}=\frac{2}{ 3}, \quad\quad
\frac{m_{2/3}}{ m_0}=\frac{1}{3}, \quad\quad
\frac{m_{3/4}}{ m_0}=0.30237... \quad,
\label{shape_sh_points}
\end{equation}
where magnetization densities are given in units of the Onsager value $m_0$.
In Fig.~\ref{fig:classical_sh_barriers} we display the length of 
classical droplet surfaces as a function of $\Omega /V$ for one, two, three 
and four corner droplets and transition points are marked by vertical lines.
To determine the energy barriers at the $2/3$ and $3/4$ transitions we 
again argue, that for the saddle point configurations one needs to deform
the $3$ corner generic droplet shape of Fig.~\ref{fig:genericb} for fixed 
volume at coexistence in such a way, that either point $d$ of the figure  
collapses 
onto point $A$, or that an additional fourth corner is included. Both of 
these shapes are contained in (\ref{drop3}) with $s=0$ for the  
volume $\Omega_{2/3}$ and $s=1$ for 
$\Omega_{3/4}$. Using these we find for the excess lengths
\ba
\frac{\partial\Omega_{2/3}-\partial\Omega_2}{4L}
&=& 0.02977... \\
\frac{\partial\Omega_{3/4}-\partial\Omega_3}{4L}
&=& 0.03441... \quad  \label{lshb}
\end{eqnarray}
with the largest barrier (\ref{lshb}) being in fair agreement with the value
(\ref{eq:r_value}) measured in the simulation.
\begin{figure}[ht]
\begin{center}
\epsfig{file=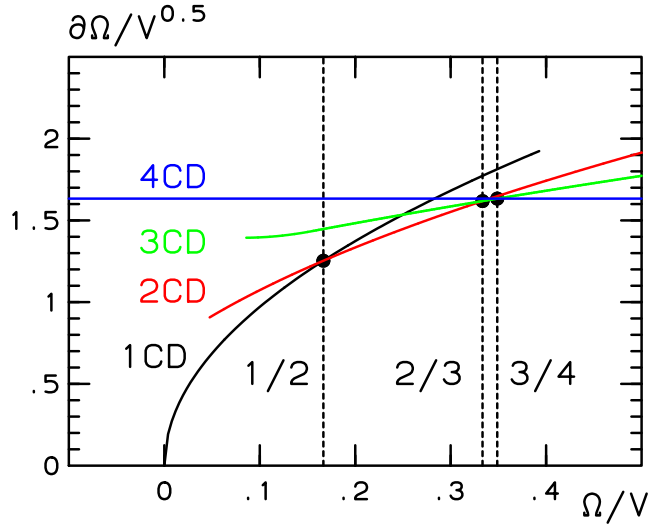,width=08cm,angle=270}
\caption{Classical droplet surfaces $\partial \Omega / V^{0.5}$ 
on a SH lattice. The curves correspond to one, two, three and 
four corner droplets with vertical lines denoting the shape 
transition points.}
\label{fig:classical_sh_barriers}
\end{center}
\end{figure}
\begin{figure}[ht]
\begin{center}
\epsfig{file=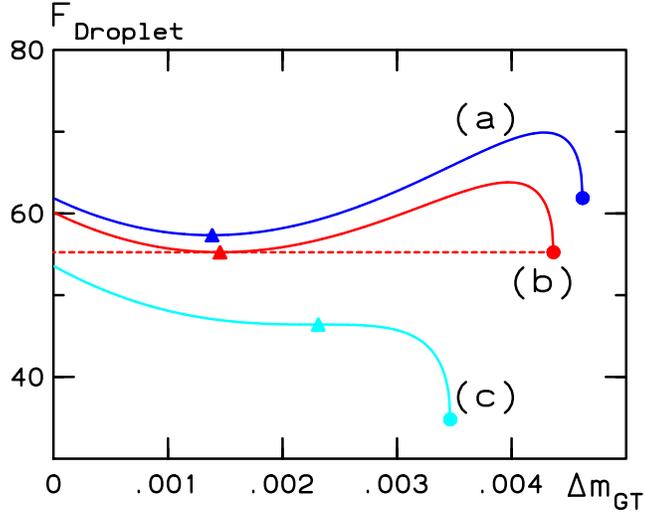,width=08cm,angle=270}
\caption{Droplet free energy $F_{droplet}$ of (\ref{dropenergy}) for three
$m$ values in vicinity and below $m_0$ as a function of the Gibbs-Thomson
shift $\Delta m_{GT}$ with $ 0 \le \Delta m_{GT} \le m_0-m $. With increasing
values of $m$ curves move from above to below. The saddle points of the 
droplet free energy are denoted by solid triangles, states with vanishing 
droplet volume correspond to solid circles. Curve (b) has a $m$-value right 
at the condensation phase transition point.}
\label{fig:gt_effect}
\end{center}
\end{figure}

\section{Finite Size Corrections}

In the limit of large box volumes $V$ and large droplets
the overall magnetization $M$ in the two phase region is created by two areas
with the Onsager value $m_0$ for the magnetization density and opposite sign:
$M=(V-\Omega)m_0-m_0\Omega$ and the droplet volume is a linear function
of the overall density
\be
\frac{\Omega}{ V}=\frac{1}{2}\left(1-\frac{m}{ m_0}\right).
\label{eq:omega}
\end{equation}
There exists a variety of finite size effects, which all contribute
to the droplet free energy and potentially modify
relation (\ref{eq:omega}) for the volume of the largest minority droplet
on finite boxes. Corrections are due to the finite curvature of the droplet 
surface or due to the restrictions posed on fluctuations by the finite size 
of the droplet or the box, or by the presence of several droplets. The 
literature hosts the following corrections to a single droplet \cite{lz90}
\begin{itemize}
\item{}Capillary wave corrections of surface string excitations with string 
       length $L$
       contribute $\ln(L)$ and constant terms to the free
       energy \cite{p88,gf88}, with presumably different coefficients
       in the strip and droplet state.
\item{}Contributions to the free energy due to degeneracies of
       strip and droplet states under translations of the lattice cubic
       group. On a finite 2D system with pbc. a strip has a degeneracy
       $2L$ and a droplet $L^2$. These entropic terms yield a contribution 
       of order $\ln(L)$ to the free energy. We expect this type
       of finite size effect to be present only at very low temperatures
       since the degeneracy is removed by thermal excitations.
\item{}Tolman corrections to the interface tension due to the small
       radius of curvature of very minute droplets \cite{t49,kb49,fw84}.
\item{}Gibbs-Thomson corrections \cite{klcs96} describing a shift of the
       bulk magnetization due to the presence of curved surfaces.
\end{itemize}
It is very instructive also, to think of $\Omega$ of (\ref{eq:omega})
as an order parameter for the formation of an minority
droplet of extensive size, which is non-zero in the two phase region
and vanishes for all states with $m > |m_0| $. There exists actually
then a phase transition associated with this order parameter, at which point
the droplet either evaporates into a gas of ``small'' droplets, or the gas 
condenses. Again for infinite systems the phase transition is located exactly 
at the Onsager $m_0$ value, but for finite systems corrections are present.
Since all the above mentioned finite size effects may conspire in
the observed shift of the transition it is an interesting and
unsettled issue to study their contributions quantitatively.
The droplet condensation phase transition has recently been studied with
Monte-Carlo simulations of Ising-models with fixed magnetization in
two \cite{ps00} and three \cite{ph01} dimensions. In these studies
the size of the largest minority cluster was measured as a function
of temperature i.e., along paths perpendicular to those we use,
measuring at fixed temperature for various values of the magnetization.
In $d=2$ the authors observed a discontinuous condensation
phase transition with finite size effects similar to those
of Fig.~\ref{fig:size} and attributed them to the Gibbs-Thomson
effect. We will quantify this statement with the following calculations
and with the fits of section \ref{condeval}. 
In three dimensions the
finite size analysis for the size of the largest minority cluster
close to the transition is complicated by the fact, that for small
enough magnetization there is a infinite percolating minority cluster
present already in the one phase region \cite{anpr77}. The precise
influence of this fact at the condensation transition deserves further
investigation. 

For the description of the condensation phase transition we assume that the 
finite box restricted $m$ partition function of a possibly discontinuous 
transition can be written as
\be\label{super_position_condensation}
Z(m,L) = e^{-F_{droplet}(m,L)} +  e^{-F_{bulk}(m,L)}  \quad ,
\end{equation}
up to corrections exponentially small in $L$ \cite{bk93}, with $F_{bulk}$ and
$F_{droplet}$ being suitable free energies in the one phase (bulk) and the two 
phase (droplet) region respectively. In the two phase region we use the 
classical theory for a single extensive droplet, as outlined in the 
preceding section, amended by 
Gibbs-Thomson (GT) corrections. The bulk phases are described by Ginzburg 
Landau (GL) theory. The Gibbs-Thomson effect \cite{klcs96} accounts for a 
finite curvature of the surface of the droplet. Microscopically one can 
explain this effect by noting, that the average coordination number of a 
spin at a surface 
with positive curvature is reduced and thereby the rate of detachment 
into the phase of opposite sign is enhanced and the other way round for 
negative curvature. To lowest order this induces a small shift $\Delta m_{GT}$ 
of the magnetization density of equal absolute value in both phases, but of 
opposite sign. Conservation of overall magnetization then leads to a shifted 
droplet volume
\be
\frac{\Omega }{ V}=
\frac{1}{ 2}\left(1-\frac{m}{ m_0}-\frac{\Delta m_{GT}}{ m_0}\right) \quad .
\end{equation}
To calculate $\Delta m_{GT}$ to lowest order 
we minimize the two phase free energy \be \label{dropenergy}
F_{droplet}=\sigma\sqrt{4\pi\Omega}+  c_2 V \Delta m_{GT}^2 \quad ,
\end{equation}
where the first part is the surface free energy of a circular droplet of 
volume (size) $\Omega$, and the second part is the excess bulk free energy 
due to the shift $\Delta m_{GT}$
in an expansion of the Ginzburg-Landau free energy up to second order 
around the bulk value $F(\pm m_0)$, which we choose to be zero for convenience. 
The coefficient $c_2=18.1252318487...$ at $\beta=0.7$ is obtained from the low
temperature series expansion results of \cite{ltres}. Minimization of 
(\ref{dropenergy}) results into
$\Delta m_{GT} \propto \partial \Omega^{-1} $ 
i.e., a shift of $\Delta m_{GT}$ proportional to the curvature of the droplet, 
which after reinsertion into the free energy leads to a curvature dependent 
downward correction to the free energy. The droplet free energy 
(\ref{dropenergy}) is displayed 
in Fig.~\ref{fig:gt_effect} for three different values of $m$ below $m_0$ 
in close vicinity of 
$m_0$ as a function of the Gibbs-Thomson shift $\Delta m_{GT}$, which for 
each $m$ value in the figure ranges in-between $\Delta m_{GT}=0$ and 
$\Delta m_{GT}=m_0-m$. With 
decreasing distance to Onsagers $m_0$ the stable saddle point solution of 
situation (a) - as depicted in the figure - turns into a metastable one (b), 
which then turns unstable in situation (c), very close to $m_0$.

In finite systems the otherwise stable classical minority droplet of the two 
phase region becomes metastable at the condensation phase transition point 
$m_{cond}(L)$, where the 
system performs a finite size rounded transition into the one 
phase region. The condensation point shift due to the finite system size
\be
\Delta m_{cond}(L)=m_0-m_{cond}(L) \quad ,
\end{equation}
corresponds to situation (b) as depicted in Fig.~\ref{fig:gt_effect} and 
can be calculated by equating the free energy
\be \label{bulkfreeenergy}
F_{bulk}= V c_2 \Delta m_{cond}^2
\end{equation}
of the bulk state - the solid circles in Fig.~\ref{fig:gt_effect} - to the 
saddle point free energy (\ref{dropenergy}) in the two phase region: the solid 
triangles in Fig.~\ref{fig:gt_effect}. Note that the surface term in 
(\ref{dropenergy}) depends 
also on $\Delta m_{cond}$. One finds the finite system condensation phase 
transition point
\be \label{shift}
\Delta m_{cond}(L)= A_{cond}  L^{-{2/3}}
\end{equation}
on toroidal $L^2$ boxes.  The exponent $-2/3$ for the finite-size behavior of the
condensation transition was already found before in the context of metastable 
decay \cite{Bin80,s80}, but the
inclusion of Gibbs Thomson corrections lead to a different value of $A_{cond}$.
The coefficient $A_{cond}$ at $\beta=0.7$ has the value 
\be \label{erg:dgtgl}
A_{cond} = m_0 \eta \frac{3}{16^{1 / 3}}  =0.23697...  \quad ,
\end{equation}
where parameter values $\eta$ and $\sigma$ are given by
\ba \label{erg:dglx}
\eta= [\frac{{\pi {\sigma}^2}}{{c_2^2 m_0^4} }]^{1 / 3} =0.20102...~\quad \\
{\sigma}  = \sqrt{W/\pi}=0.90358... \quad . 
\ea
The droplet size $\Omega_{cxc}(L)$ at the condensation phase 
transition is
\begin{equation}\label{dropsize}
\Omega_{cxc}(L)= \frac{V }{ 3}  
\frac{ \Delta m_{cond}(L) }{ m_0 }=0.07977...~ L^{{ 4 / 3}} \quad ,
\end{equation} 
and one can also calculate the nucleation barrier $B_{nucl}$, which at the 
condensation phase transition corresponds to the maximal excess droplet free 
energy in-between states at the saddle point and states at 
$\Delta m_{GT}=m_0-m$. One finds
\begin{equation}\label{thebarrier}
\frac{B_{nucl} }{ \sigma\sqrt{4\pi\Omega_{cxc}} } =0.174038...\quad 
\end{equation} 
in units of the coexisting droplets surface free energy and 
$B_{nucl}/F_{droplet} = 0.154701$ in units of the total free 
energy of the coexisting droplet.
The Gibbs-Thomson corrected droplet looses its stability at the point
\ba \label{erg:stability}
\Delta m_{unstable}(L) 
  = m_0  \eta   6^{1/3} (\frac{3 }{8})^{2 / 3} ~ L^{-{ 2 / 3}} \quad  \\
                   =0.18808...~ L^{-{ 2 / 3}} \quad ,
\ea
which inside of the metastable region terminates the droplet metastable branch. 
Condensation phase transition point scaling with the power law $L^{-{2/3}}$ can 
also be proven with rigorous methods (see \cite{biv00}
for a recent overview). Furthermore one finds at the condensation phase 
transition point the linear relation
\be \label{gt:shift}
\Delta m_{GT}(L)=\frac{1 }{ 3}\Delta m_{cond}(L) \quad ,
\end{equation}
which expresses the Gibbs-Thomson shift in units of the finite size location 
of the phase transition point. This linear relation is especially interesting 
as the ratio
\begin{equation}\label{qvalue}
Q= \frac{  {V  \Delta m_{cond}(L)  / 2 m_0 } }{ \Omega_{cxc}(L)} 
\end{equation}
is predicted to have the value $Q=3/2$. If one 
neglects Gibbs-Thomson corrections altogether one finds $Q=1$. The value of  
$Q$ does not 
depend on the
particular values for the surface free energy, nor on the parameter 
value $c_2$ (which all factor out in 
the calculation). A stringent test on the presence of Gibbs-Thomson corrections 
thus can be devised, if in the Monte-Carlo simulation one 
measures both, the coexisting droplets size $\Omega_{cxc}(L)$ and 
location of the condensation phase transition $\Delta m_{cond}(L)$.

\section{SIMULATION RESULTS}

We have simulated the 2D Ising-model at $\beta=0.7$ ($k_BT=1.428\dots$) 
on toroidal lattices and on cube-surface $SH(L)$ lattices.
In one set of Muca simulations we cover all states from the strip to
the bulk phase and study $L=12$ up to $L=44$ toroidal $L^2$
boxes and cube-surface lattices of parameter values $L=4$
up to $L=26$. In another set of simulations the condensation point phase 
transition is studied on toroidal lattices of sizes $40^2$ up to $400^2$, 
again also with the use of Muca simulations covering however a smaller
$m$-interval in vicinity of the condensation phase transition.

\subsection{SHAPE TRANSITIONS}

In order to study the droplet shape transitions on the torus and on 
SH lattices in detail we consider the constraint magnetization ensemble of 
the Ising-model. Its partition function $Z(m,L)$ is
\be \label{respart}
Z(m,L) = \sum_{conf.}e^{ -\beta H}
\delta(m-\frac{1 }{ V}\sum_{i}s_i)
\end{equation}
and we determine the expectation values of observables at fixed magnetization 
$m$. Muca ensemble simulations are perfectly suited for the evaluation of 
restricted expectation values. All one has to do is to average over operator 
values at $m$, if 
the magnetization $m$ is visited in the Muca simulation. As we are dealing 
with droplets it is sensible to introduce the $m$-dependent functions
\ba
\partial \Omega_0(m)&=& \sqrt{2 \pi V(1 -\frac {m}{m_0})} \\
\Omega_0(m)&=& \frac{V }{ 2}(1 - \frac{m}{ m_0}) \quad ,
\end{eqnarray}
which for large system volume $V$ denote the classical surface and volume of a 
circular droplet at fixed overall magnetization $m$. For each configuration at 
$m$ we evaluate connectivity components i.e., clusters of spins. Nearest 
neighbor spins belong to the same cluster, if they have the same value. The 
size $\Omega$ - the number of sites - is determined for each cluster and 
clusters are sorted with respect 
to their size. The second largest cluster defines the object of interest and 
corresponds to the minority phase droplet. Its expectation value in the 
magnetization bin $m$ is denoted $<\!\Omega\!>(m)$, the ``volume'' of the 
minority phase droplet. It should 
be noted that $<\!\Omega\!>(m)$ not necessarily has to agree with 
$\Omega_0(m)$, if finite size effects are present.
\begin{figure}[tb]
\centerline{ \begin{minipage}[c]{13.2cm}
(a) \psfig{file=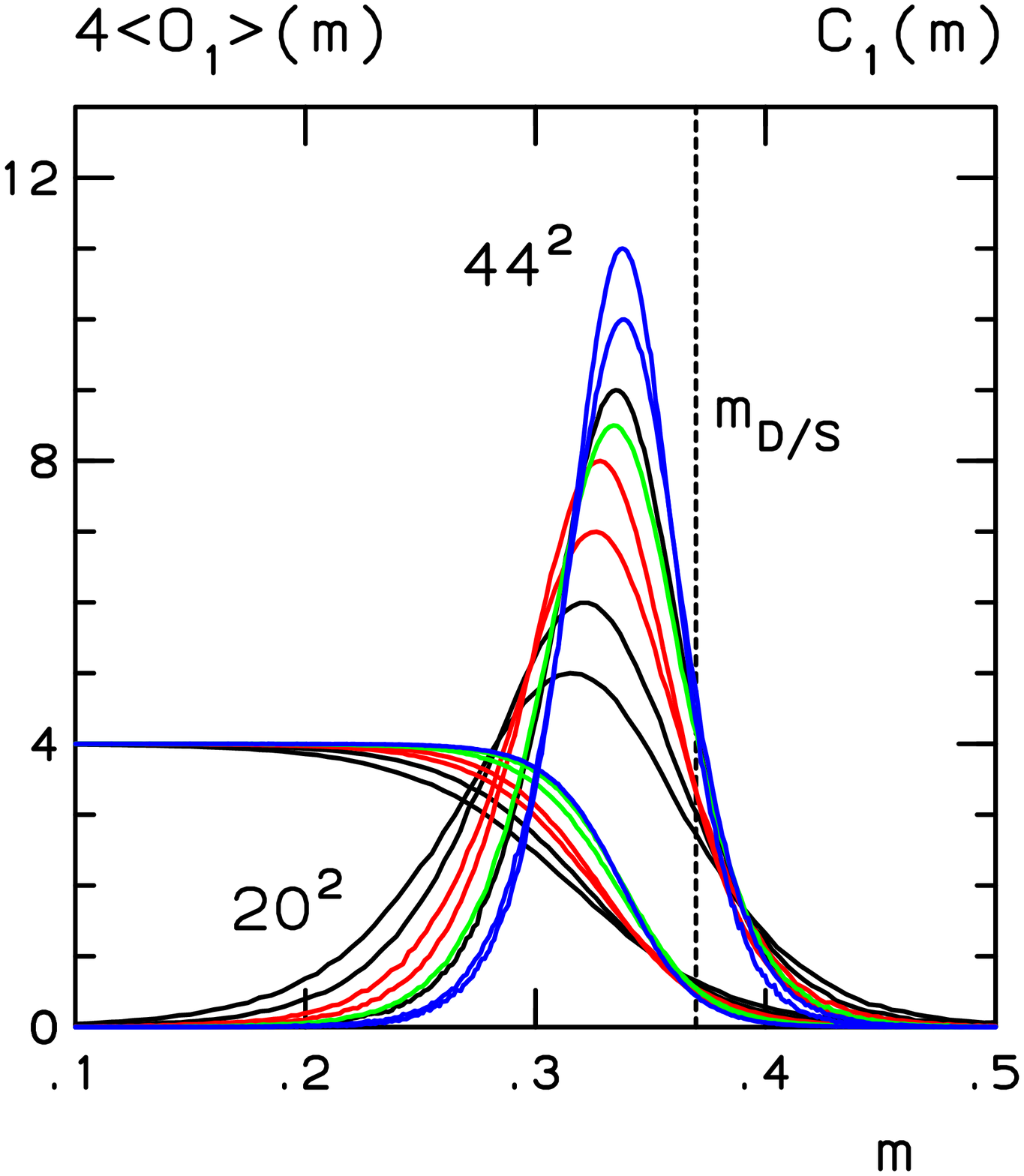,angle=000,width=5.5cm}
(b) \psfig{file=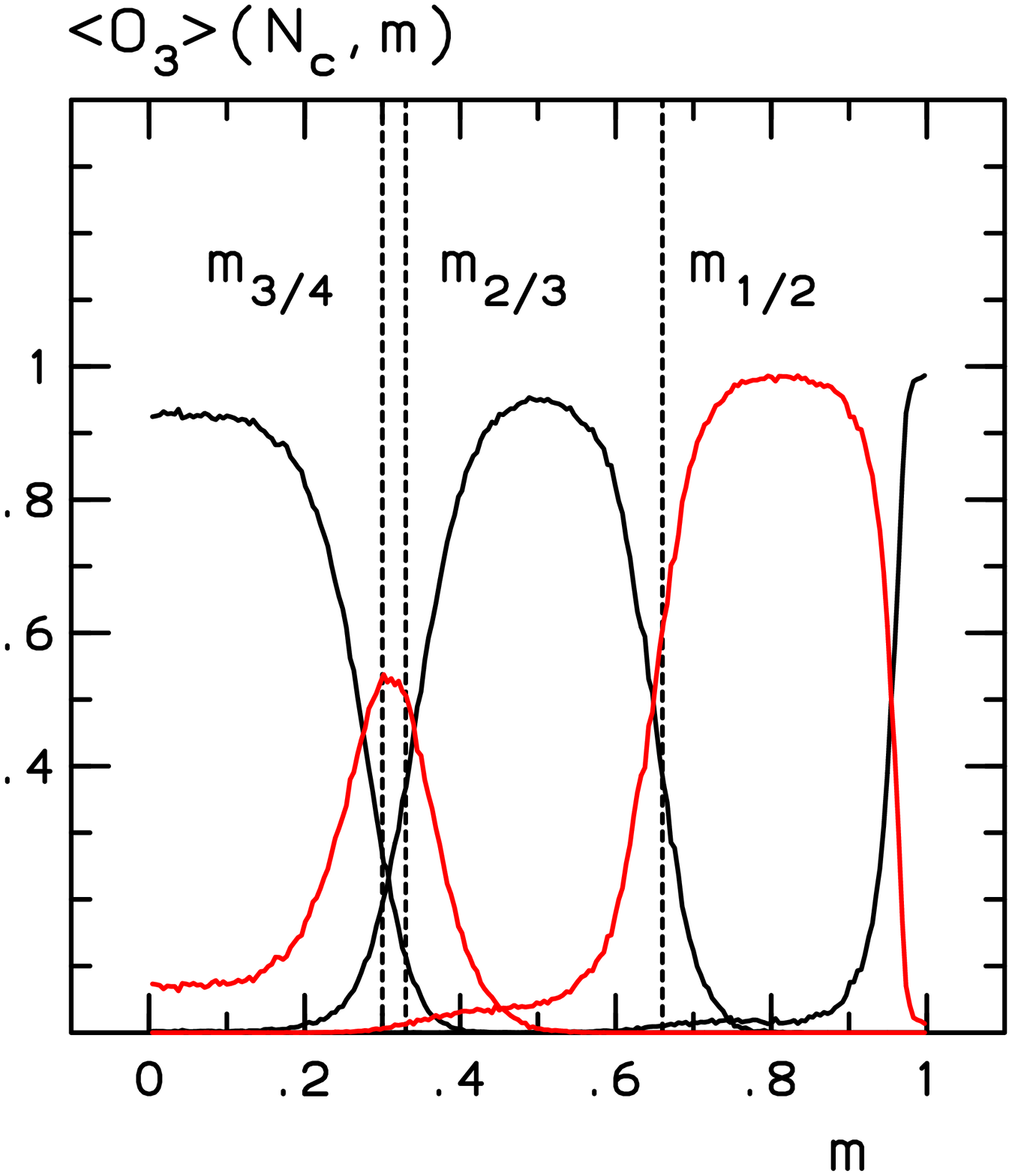,angle=000,width=5.5cm}
    \end{minipage}}
\caption[a]{ (a): Droplet to strip transition geometric order parameter 
             $4<\!O_{1}(m)\!>$ and susceptibility $C_{1}(m)$ for the torus 
             with pbc.
             and (b): geometric order parameter $<\!O_{3}(m,N_c)\!>$ on a 
             $SH(26)$ lattice with $N_c=0,1,2,3$ and $N_c=4$. The $5$ different 
             $N_c$-dependent curves have maxima, which from the right to the 
             left 
             correspond to values $N_c=0,1,2,3$ and to $N_c=4$. Vertical lines 
             denote classical 
             shape transition points. 
           }
\label{fig:percolation_order-parameter}
\end{figure}
For the toroidal lattice geometry (pbc.) and for the minority phase droplet
a rectangular bounding box with linear sizes $L_1$ and $L_2$ is determined in 
such a way, that the droplet exactly fits into the box. From the geometric 
numbers $L_1$ and $L_2$ two geometric order parameters are formed, namely
\begin{eqnarray}
      O_{1}(m)&=&   \delta ( L-{\rm max}(L_1,L_2) )\\
      O_{2}(m)&=&   \frac{ {{\rm max}(L_1,L_2)} }{ {L} } \quad .
\end{eqnarray}
Values for the operator $O_{1}(m)$ are zero and unity. At the position 
$m_{D/S}$ of the droplet strip transition we expect to find a finite size 
rounded 
jump of the expectation value $<\!O_{1} (m)\!>$ from value zero to unity. 
The order parameter $O_{2}(m)$ on the other hand has several discrete values 
in the interval $[0,1]$. One can construct a probability distribution 
function $P(O_{2})$ for the occurrence of values $O_{2}$ in the restricted 
partition function (\ref{respart}) at $m$. If our classical arguments on the 
nature of the droplet to 
strip transition are correct, then  one expects to find a double peaked 
distribution function $P(O_{2})$ in the vicinity of $m_{D/S}$ with a barrier, 
which is related to the excess length. Susceptibilities $C_{1}(m)$ and 
$C_{2}(m)$ are defined by
\be
C_{1}(m)=L<\!( O_{1}(m)-<\!O_{1}(m)\!>)^2\!>
\end{equation}
and similar for $C_{2}(m)$ with $O_2$, where $<\, >$ again denotes the 
expectation value at 
given $m$. These susceptibilities show a finite size rounded peak at the 
shape transition defining finite size shifted $m_{D/S}(L)$-values. 

For the minority droplet on SH lattices we define a geometric order parameter 
sensitive to the number $n_c$ of corners occupied by the
droplet. At magnetization $m$ we count the number $n_c$ and define 
\be
O_{3} (m,N_c)=   \delta ( N_c-n_c ) \quad ,
\end{equation}
which at given $m$ receives unity contributions only if $N_c$ corners are
occupied. We expect e.g., that expectation values  $<\!O_{3} (m,2)\!>$
yield non-vanishing values $<\!O_{3} (m,2)\!> \approx 1$ only, if
the magnetization density $m$ lies in-between the $1/2$ and $2/3$ shape 
transition magnetization density values.
Data for the geometric order parameters $<\!O_{1}\!>(m)$ 
on toroidal lattices and data for $<O_{3} (m,N_c)>$ with $N_c=0,...,4$ on a 
$SH(26)$ 
lattice are displayed in Figs. \ref{fig:percolation_order-parameter}a and 
\ref{fig:percolation_order-parameter}b.
\begin{figure}[tb]
\centerline{ \begin{minipage}[c]{13.2cm}
(a) \psfig{file=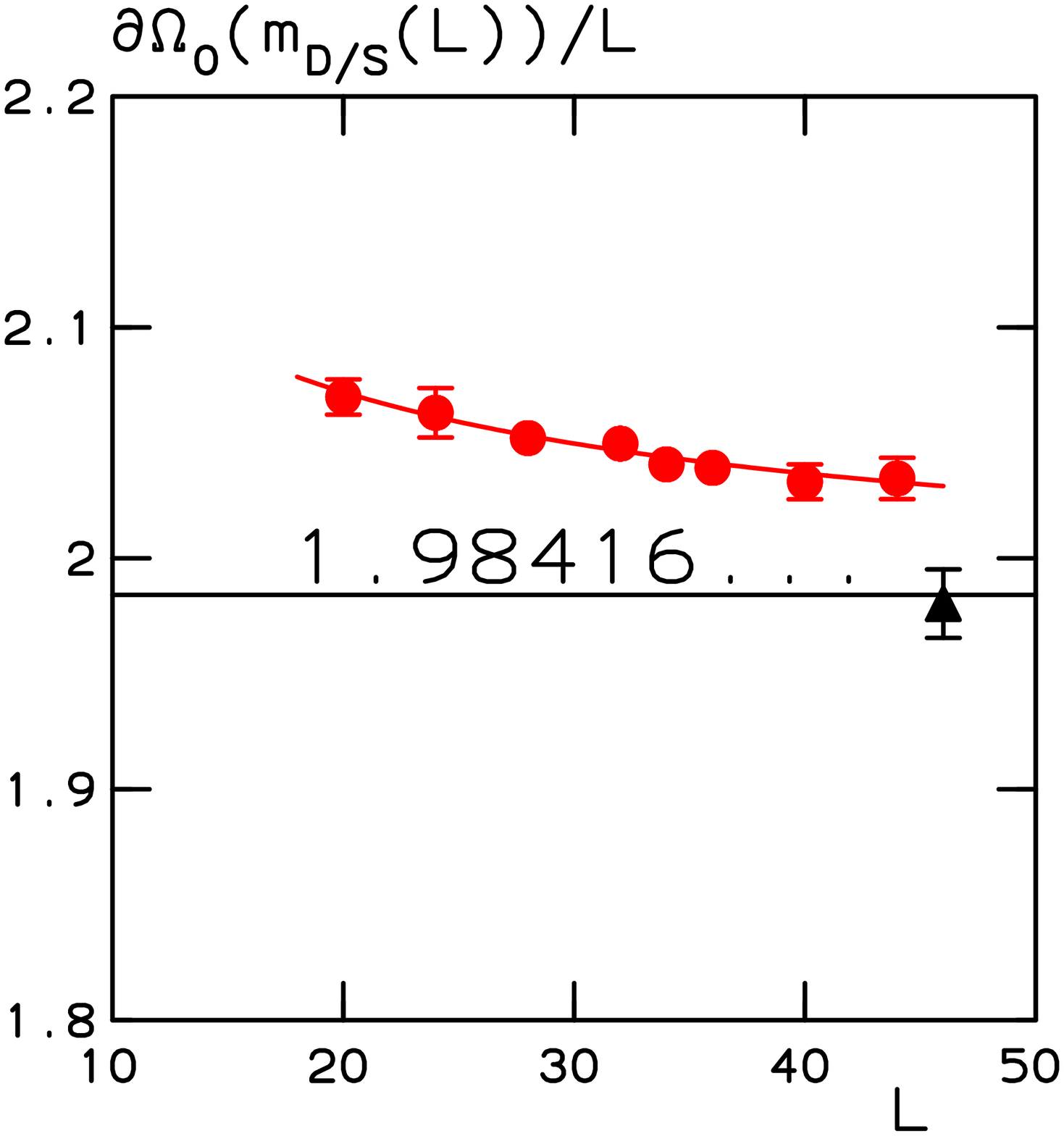,angle=000,width=5.5cm}
(b) \psfig{file=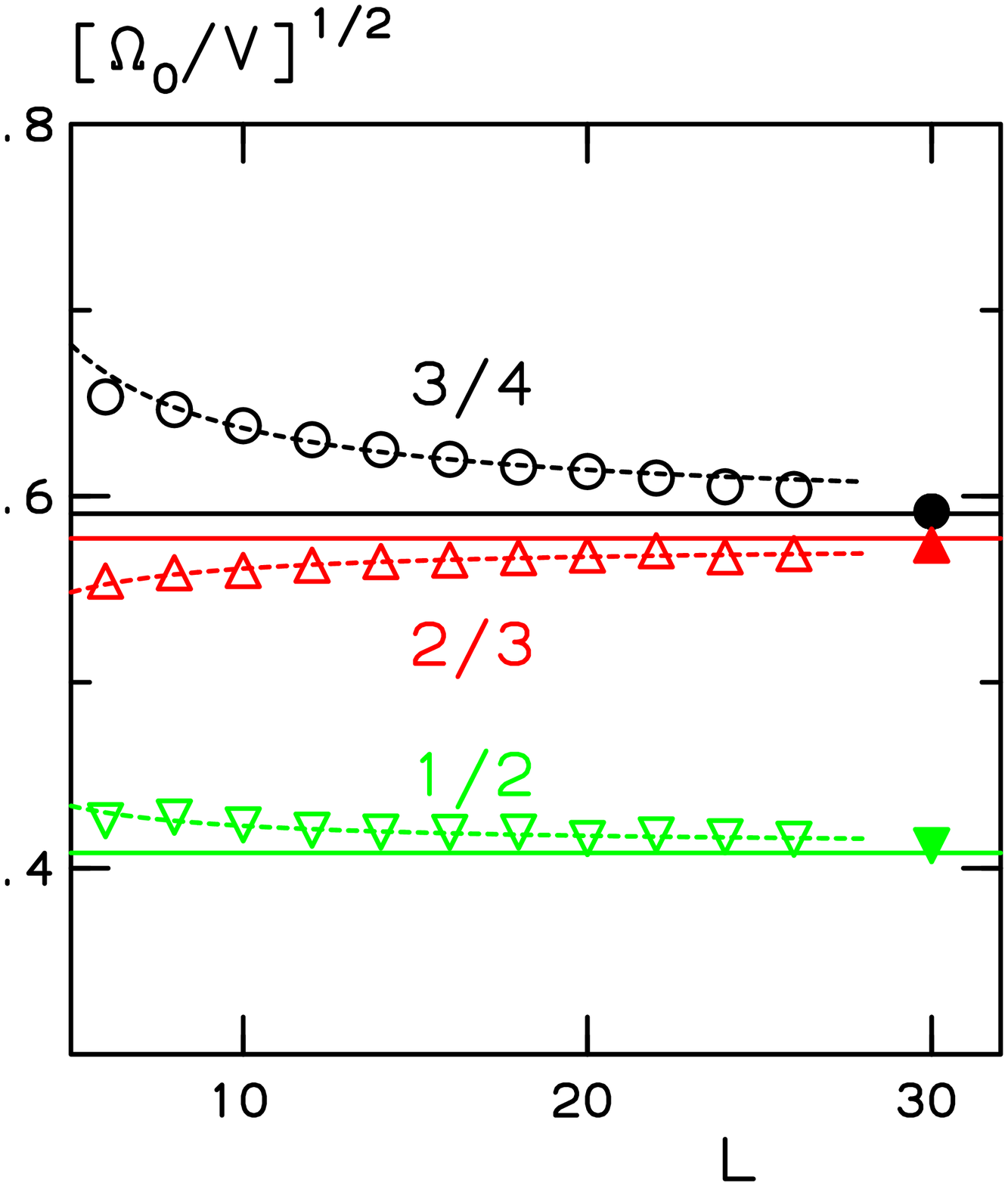,angle=000,width=5.5cm}
    \end{minipage}}
\caption[a]{ (a): Fit to $m_{D/S}$ on the torus. The function 
             $\partial\Omega_0/L$
             is evaluated at the measured finite size shifted values of  
$m_{D/S}(L)$
             and plotted as a function of $L$. The line corresponds to
             the exact result, while the solid triangle corresponds to the
             infinite volume extrapolation. The curve corresponds to the fit 
             as explained 
             in the text. (b): Fit to the three shape transition
             points on $SH(L)$ lattices. Here the function $\sqrt{\Omega_0/V}$
             is evaluated at the measured finite size shifted magnetization  
density
             values and plotted as a function of $L$. Again horizontal lines
             correspond to droplet calculations, while curves correspond to fits
             explained in the text. Solid symbols correspond to the infinite
             volume extrapolation.
           }
\label{fig:position}
\end{figure}
\begin{figure}[tb]
\centerline{ \begin{minipage}[c]{13.2cm}
(a) \psfig{file=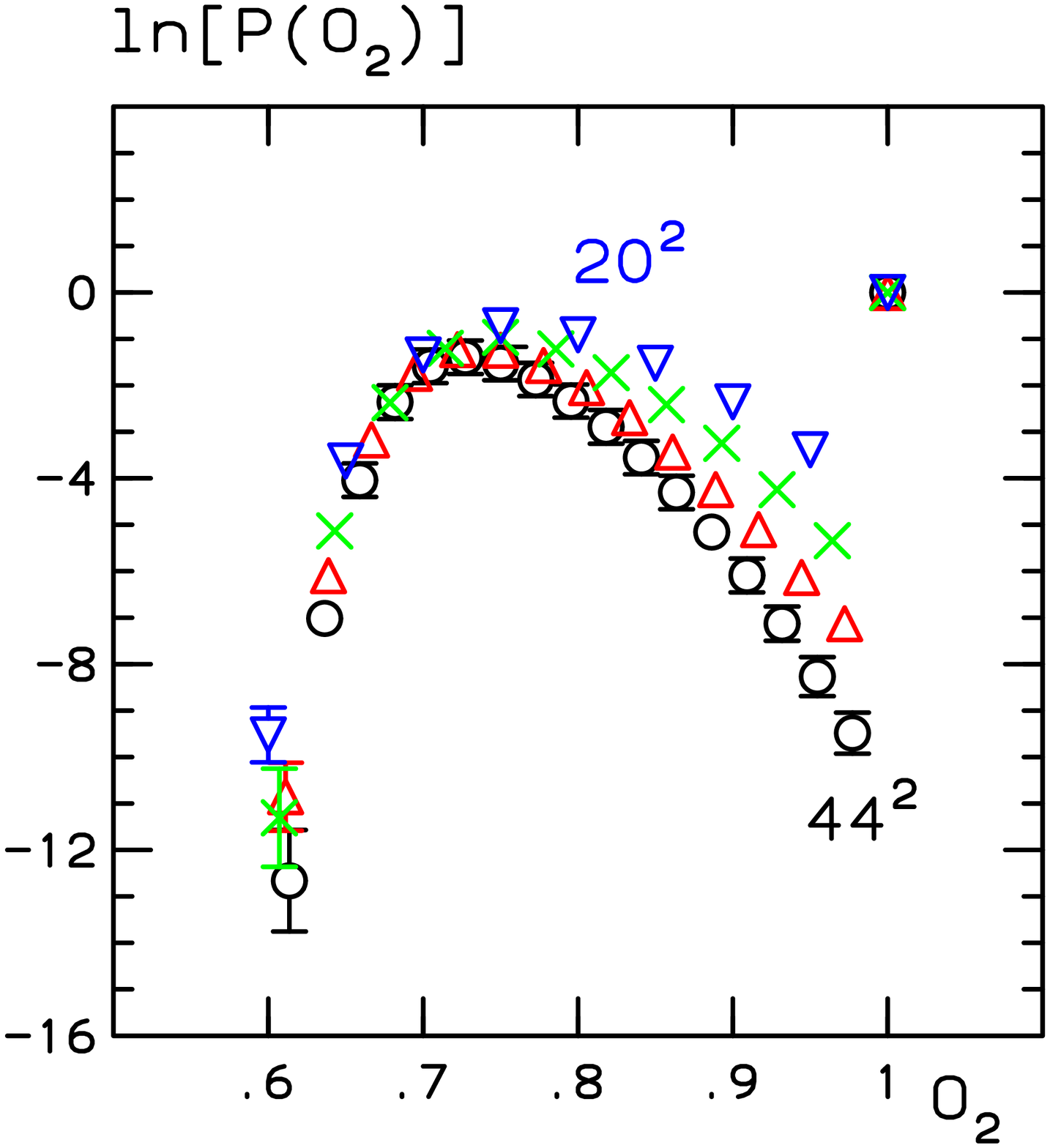,angle=000,width=5.5cm}
(b) \psfig{file=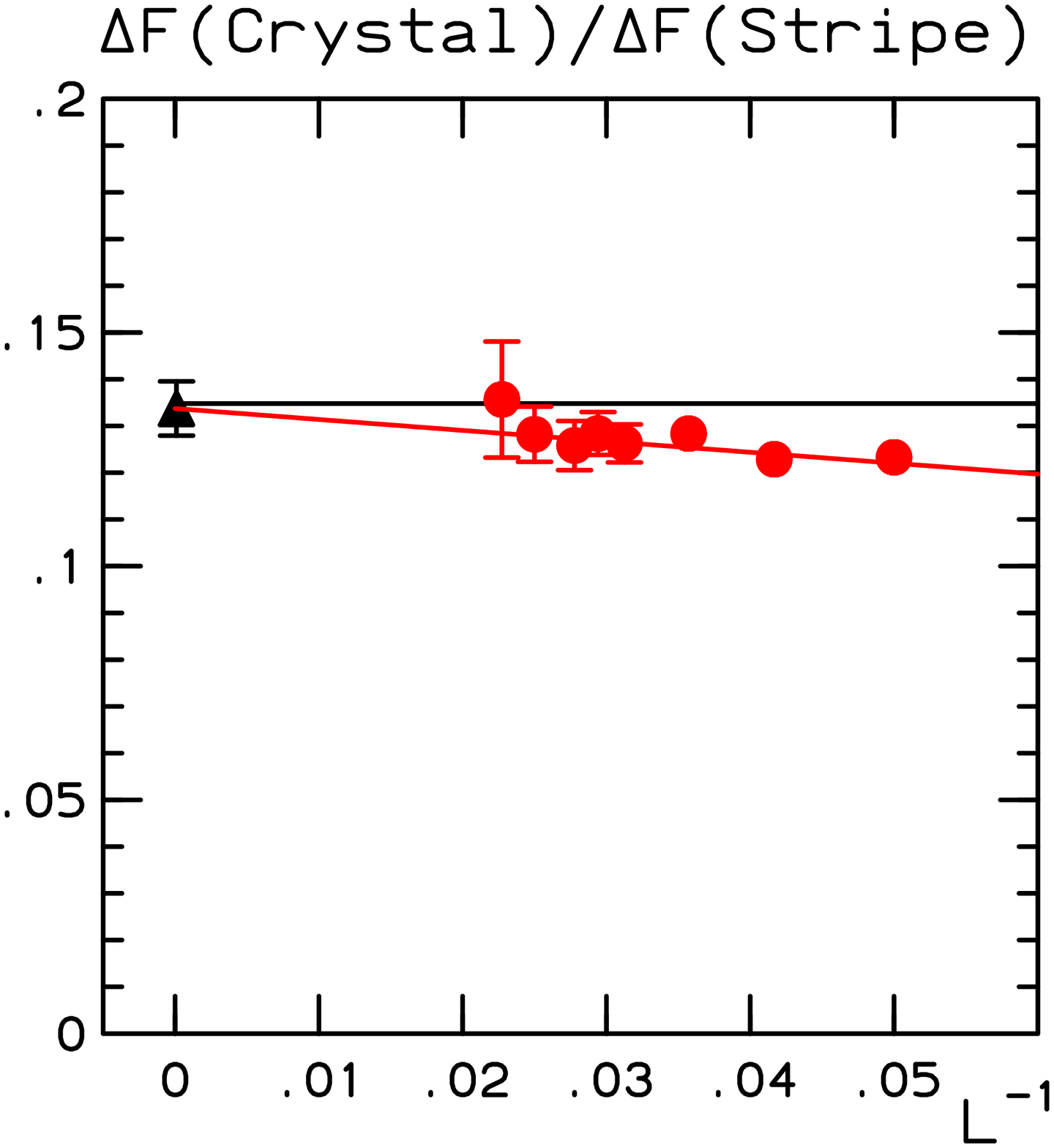,angle=000,width=5.5cm}
    \end{minipage}}
\caption[a]{
(a) Geometric order parameter $O_2$ distribution on the torus in vicinity 
of $m_{D/S}$ and (b) barrier values as a function of $1/L$. The horizontal line
denotes the classical droplet theory result, the solid triangle the infinite 
volume extrapolation.}\label{fig:distri}
\end{figure}
The maximum positions of the susceptibility in
Fig.~\ref{fig:percolation_order-parameter}a on toroidal lattices define finite 
volume estimates of the magnetization density $m_{D/S}$ at coexistence.  
Finite size 
correction are not particularly small and thus a careful infinite volume 
extrapolation 
is needed. For every pair of values $L$ and $m_{D/S}(L)$ we calculate 
$\partial \Omega_0(m_{D/S}(L))/L$ and with the use of the extrapolation
\be
\partial \Omega_0(m_{D/S}(L))/L=\sqrt{2 \pi (1 - \frac{ m_{D/S}}{ m_0})}
+A_{D/S} \frac{{\rm ln}(L)}{ L}
\label{eq:fssmppt}
\end{equation}
$m_{D/S}$ is determined. We mention, that ${{\rm ln}(L) / L}$ 
finite size corrections are predicted by capillary wave fluctuation 
corrections to the droplet free energy \cite{p88,gf88} and by entropic terms. 
The fit in accord with (\ref{eq:fssmppt}) 
has a $\chi^2_{dof}$ value $0.32$ and results into a transition point at 
$m_{D/S} = 0.376(9)$, which within error bars coincides with the exact 
result (\ref{strip_droplet_point}). The data and the fit are displayed in 
Fig.~\ref{fig:position}a. A similar analysis based on $C_{2}(m)$ data yields 
the determination $m_{D/S} = 0.360(10)$, which again agrees with the exact 
result. Similar as in torus case we also calculate susceptibilities of the 
$O_3$ order parameter on $SH(L)$ lattices and determine finite size shifted shape 
transition points. The results are displayed in Fig.~\ref{fig:position}b 
as a function of $L$ on
$SH(L)$ lattices. In this case three different $i=1,2,3$ infinite volume 
transition points are determined via the extrapolation 
\be
\sqrt{ \frac{ \Omega_0(m_{i/i+1}(L))}{V} }=\sqrt{ \frac{1}{2} 
(1 - \frac{ m_{i/i+1}}{ m_0})}
+B_{i/i+1} \frac{ 1}{ L} \quad . 
\end{equation}
We did not include a ${{\rm ln}(L) / L}$ term in this Ansatz, since capillary 
wave corrections are the same on both sides of the transition and entropic 
terms should be absent since the droplets center of mass is energetically 
pinned. 
We obtain the values $m_{1/2}=0.661(2)~m_0$, $m_{2/3}=0.342(3)~m_0$ 
and $m_{3/4}=0.305(3)~m_0$, which within a systematic one percent relative 
error all agree with the values of (\ref{shape_sh_points}). The small discrepancy 
is presumably caused by the isotropic surface free energy of our (approximative) 
droplet calculation for $SH$ lattices.
The susceptibility $C_{2}(m)$ on toroidal lattices has its maximum value at 
positions $m_{D/S}(L)$ and we can determine the probability distribution 
$P( O_{2})$ of the order parameter $O_{2}$ there. The data are displayed in 
Fig.~\ref{fig:distri}a 
for $L=20,28,36$ and $L=44$ lattices. One observes clear double peaks 
with the peak for the strip concentrated in the single point at $O_2=1$.
In-between one finds states, which with increasing lattice size
become less an less probable. These suppressed states are the saddle point 
crystal shapes 
of Fig.~\ref{fig:shapes}. A split of phase space into two disconnected 
regions with a free energy barrier in-between proves the discontinuous nature 
of the droplet strip transition and 
one can determine the barrier height. The distribution functions have two 
maxima values $P_{max,1}$ and $P_{max,2}$ and a minimum value $P_{min}$. We 
form
\be
\Delta F_{droplet}=\frac{1}{2}\ln (P_{max,1}P_{max,2}/P_{min}^2) \quad ,
\end{equation}
which we measure in units of
$
\Delta F_{strip}=2L \sigma_{0}
$
and the result is displayed in Fig.~\ref{fig:distri}b as a function
of $1/L$. A linear fit in $L^{-1}$ (the curve in the figure) results into the
value
\be
\Delta F_{droplet}/ \Delta F_{strip} = 0.133(6)
\label{crystal}
\end{equation}
at a $\chi^2_{dof}$ value $0.40$ for the fit. The free energy barrier measured 
in exponential slowing down of Muca simulation (\ref{bsds}) agrees with the 
free energy barrier measured in the suppression of saddle point crystal shapes 
(\ref{crystal})
and with the result of classical droplet theory (\ref{bsd}).
\begin{figure}[t]
\begin{center}
\epsfig{file=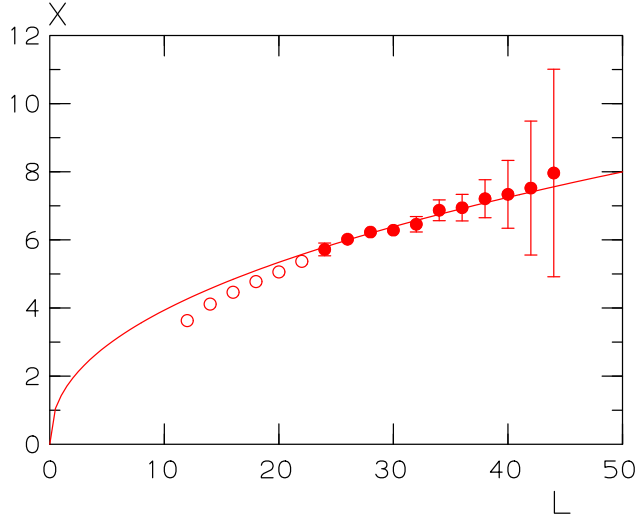,width=8cm,angle=270}
\caption{
$X$ values as defined in (\ref{x_values}) as a function of $L$ and fit 
as explained in the text.}
\label{fig:x}
\end{center}
\end{figure}

\subsection{MORE DROPLET FREE ENERGY CORRECTIONS}\label{toreval}

A direct consequence of a discontinuous behavior of the restricted 
partition function $Z(m,L)$ at the droplet strip shape transition $m_{D/S}$ 
on the torus is again the validity of a superposition Ansatz \cite{bk93}
\be\label{d__s_part_func}
Z(m,L)= e^{-F_{strip}(m,L)} + e^{-F_{droplet}(m,L)} \quad ,
\end{equation}
which we use to determine free energy corrections. If we define 
the quantity 
\be \label{x_values}
X=\frac{Z(m=m_{D/S},L)}{Z(m=0,L) }-1  
\end{equation}
we observe, that classical bulk and surface free energy contributions 
proportional to $L^2$ and $L$ cancel in 
$F_{strip}(m=0,L)-F_{droplet}(m=m_{D/S},L)$. Note that the classical surface
contributions to the free energy of a strip state at $m=0$ equal 
those at $m=m_{D/S}$ and in addition classical surface portions of $F_{strip}$ 
equal those of $F_{droplet}$ at $m=m_{D/S}$. We obtain the representation
\be
X = e^{G_{strip}(m=0,L)-G_{droplet}(m=m_{D/S},L)}
\end{equation}
where the $L$-dependence of the function $G$ at most is of the order $o(L)$
and possibly contains the whole set of finite size corrections 
to strip and droplet states.

One of the quantities, which in the Muca simulation is determined, is the 
magnetization probability $P_L(M)$. It counts the probability to find 
magnetization $M$ in the unconstrained Ising model and is proportional
to the constraint partition function $P_L(M=mV) \propto Z(m,L)$. The value 
$P_L(M=0)$
is easily measured at the magnetization bin $M=0$, while $Z(m=m_{D/S},L)$ with
$m_{D/S}=0.36974...$ is interpolated from data at magnetization bins closest to
the value $Vm_{D/S}$. Fig.~\ref{fig:x} displays the data for $X$ as a function 
of $L$. Right at the crystal shape transition and with increasing lattice size
droplet states relative to strips turn out to be more and more probable.
For large systems the data can be fitted with a power law in $L$ 
(the solid symbols in the figure)
\be
X=A_X L^\alpha
\end{equation}
and at a $\chi^2_{dof}$ value of $0.12$ we obtain with $A_X=1.4(4)$
an exponent value of $\alpha=0.44(8)$.  Such finite size effects do not have 
their origin in Gibbs-Thomson corrections but either are 
generated by capillary wave fluctuation corrections - or by the count of strip 
and droplet states with respect to translations. At zero temperature each 
translational degree of freedom reduces the free energy by a 
term $-\ln(L)$ and for finite temperature, each fluctuating surface contributes 
$+\frac{1}{2}\ln(L)$ via the capillary wave expansion. Our theoretical 
prediction for 
$\alpha$ therefore is $\alpha=-1+2(\frac{1}{2})+2-\frac{1}{2}=3/2$, if 
translational degeneracies are counted - or 
$\alpha=2(\frac{1}{2})-\frac{1}{2}=1/2$, if 
temperature lifts degeneracies. The measured value
is consistent with one half and thus temperature is too high for a ``naive'' 
zero temperature 
count of degeneracies. Note, that two independent surface strings were assumed 
for the strip and one for the droplet.
\begin{figure}[tb]
\centerline{ \begin{minipage}[c]{13.2cm}
(a) \psfig{file=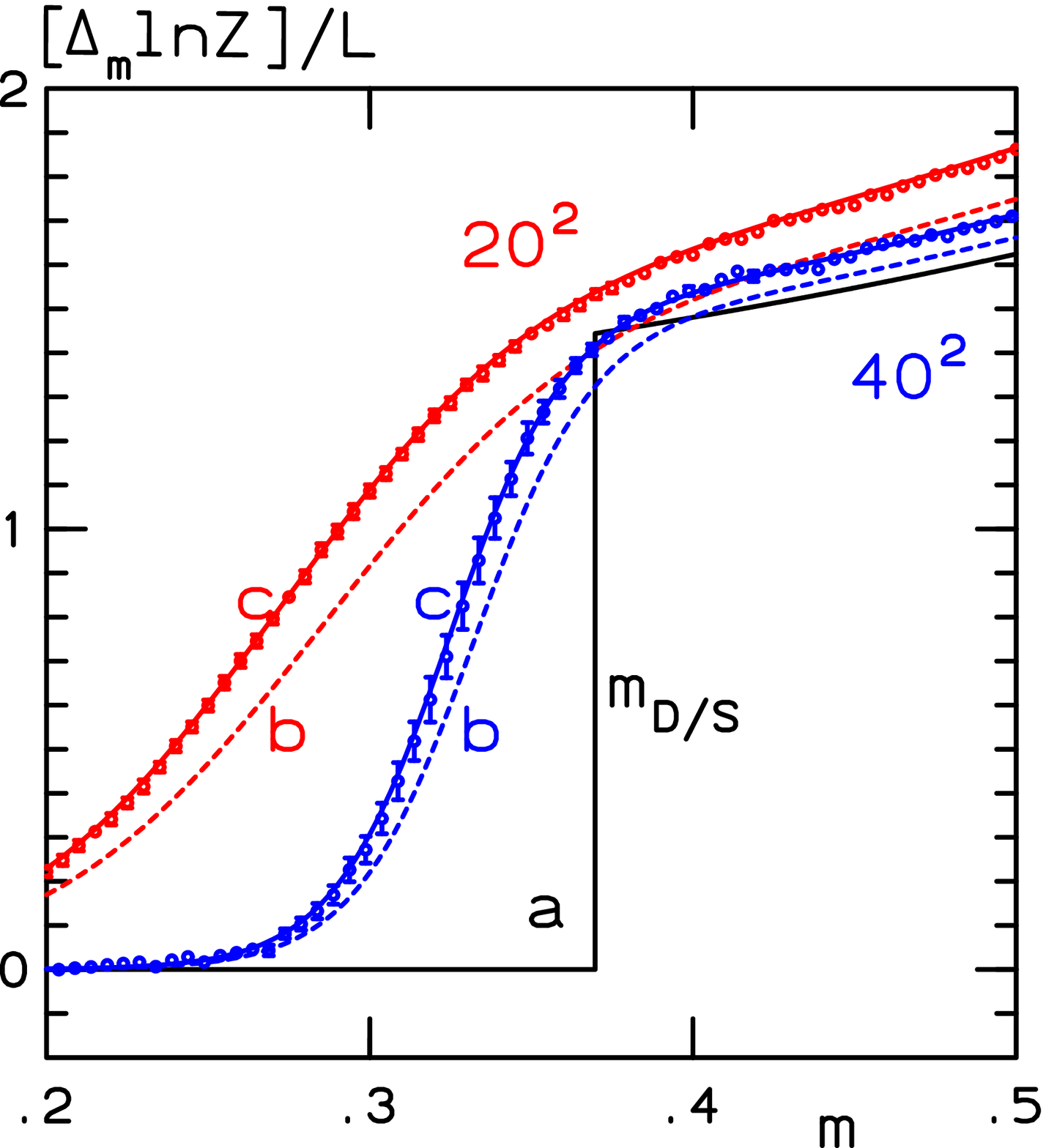,angle=000,width=5.5cm}
(b) \psfig{file=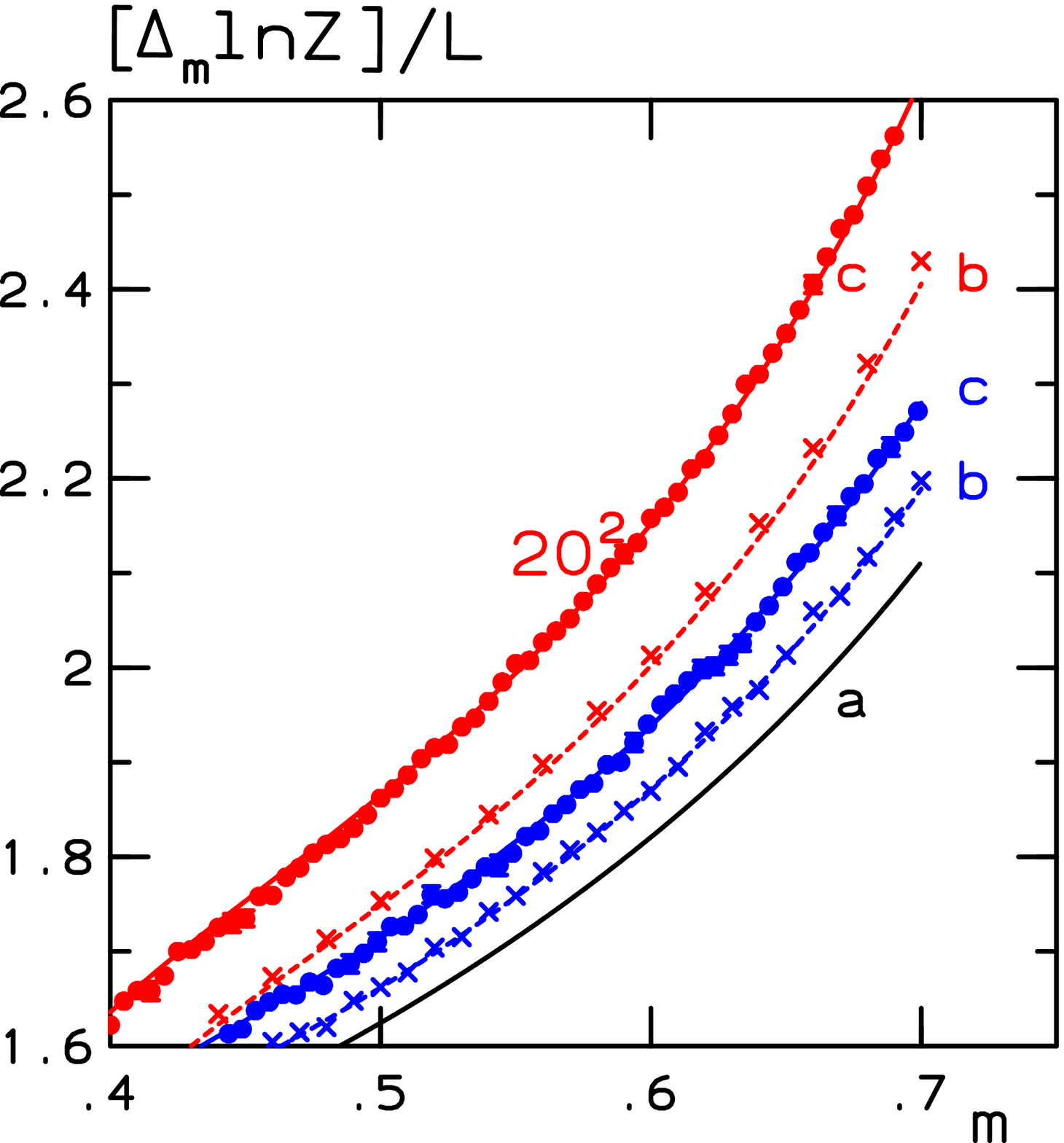,angle=000,width=5.5cm}
    \end{minipage}}
\caption[a]{Finite size rounding of the  discrete derivative $(\Delta_m\ln Z)/L$
            in vicinity 
            of the
            crystal shape transition (a) and in the droplet phase (b).
            For reasons of clarity we only present data for two lattice sizes, 
            $20^2$
            and $40^2$. The curves labeled ``c'' in the figures exactly match 
            the measured
            data (circles) and correspond to theoretical predictions for  
droplet and 
            strip free
            energies. The data denoted by crosses in (b) are results from a  
simulation
            with a modified partition function (one spin is fixed to $-1$) and
            agree with curves labeled ``b''. These are based on the droplet  
free energy without
            logarithmic phase space factor in the droplet volume.            
           }
\label{fig:z}
\end{figure}
To probe free energies further, we introduce the  discrete 
$m$ partition function derivative
\be  \label{deltaz}
\Delta_m {\rm ln}Z
:=    
\frac{ 
{\rm ln} Z(m+\Delta_M/V,L)- {\rm ln}Z(m,L)   }{  (\Delta_M/V) 
}  \quad ,
\end{equation}
which for small $\Delta_M$ is proportional to $\frac{\partial F(m)}{\partial  
m}|_{T,V}$.
Thus $\Delta_m {\rm ln}Z/L^d$ can be interpreted as a magnetic field  
\cite{lnr95} or as
a chemical potential \cite{fb82} in the lattice gas interpretation.
We have chosen a suitable value $\Delta_M=20$ and the measured Monte Carlo data 
are displayed in Figs. \ref{fig:z}a and  \ref{fig:z}b for $m$-values
in vicinity of $m=m_{D/S}$ in (a) and for $m$ values in the droplet phase in 
(b).  Similar data for the 2D Ising model at temperature $T=0.8T_c$,  
presented in Fig.~10
of \cite{lnr95}, also showed the presence of the strongly rounded stripe droplet 
transition, without attempting a quantitative fit of the data.
For reasons of clarity we only present results for two lattices,  
$20^2$ and  $40^2$ in size. The data, the circles in the figures, can 
be compared to several finite size rounding predictions 
(the curves in the figures), which are based on the 
superposition Ansatz  (\ref{d__s_part_func}). The infinite volume 
prediction, which is labeled ``a'', jumps at $m_{D/S}$ from the value zero 
(strip phase) to a finite value (gap) and in the droplet phase follows the 
classical result (\ref{exact1}). The curves labeled ``b'' in the figure 
correspond
to finite size rounding in accord with (\ref{d__s_part_func}) and with free 
energy functions
\ba
F_{strip}= 2\sigma_{0} L + {\rm ln}(L) \\
F_{droplet}=\sigma\sqrt{4\pi\Omega}+  c_2 V \Delta m_{GT}^2
+\frac{1}{ 2}{\rm ln}(\partial \Omega )
\end{eqnarray}
including classical terms as well as capillary wave fluctuations corrections.
As can be noted, the presence of further correction terms is suggested.

The translational invariance of a droplet of $-$ spins 
floating in a background of $+$ spins can easily be broken, if one spin of the 
partition 
function $Z_{droplet}$ is fixed to the value $-1$, which in consequence
lowers the partition function by a factor $f$ : $Z_{fixed~spin}=fZ_{droplet}$
with $f<1$. At zero temperature one finds $f = \Omega /V$ and thus a 
``microcanonical droplet phase space volume'' correction to the droplet free 
energy of the form
\ba\label{drop_new}
\hat{F}_{droplet}=F_{droplet}+ {\rm ln}( \frac{\Omega}{V} ) \quad ,
\end{eqnarray}
quite similar to the microcanonical phase space volume of gases, is predicted.
Adding this term to the droplet free energy in the droplet phase moves curves 
labeled ``b'' to curves labeled ``c'' in Fig. ~\ref{fig:z}b, which exactly 
reproduce the data. One can also do a simulation in the modified theory, which
differs from the original one by the fixation of one arbitrary single spin to
the value $-1$, which  never is updated. The crosses of Fig.~\ref{fig:z}b
correspond to data from such a simulation and as can be seen: they come to lie
on the curves labeled ``b'', the free energy form without logarithmic phase space
factor in the droplet volume.

It is interesting to ask the non-trivial question, whether similar phase space 
corrections contribute to the strip free energy. Strips are separated from 
droplets by barriers and thus the full occupancy of phase space (with logarithmic 
phase space factor in the droplet volume) may never be reached. A situation like 
this can be termed: spontaneous breaking of translational invariance 
for strip states, which according to our
findings actually is realized in the $\beta=0.7$ 2D Ising model. The curves 
labeled (c) of Fig.~\ref{fig:z}a have been calculated with the droplet free  
energy 
$\hat{F}_{droplet}$ 
and with
\ba\label{strip_new}
\hat{F}_{strip}=F_{strip}- {\rm ln}(2) \quad 
\end{eqnarray}
and coincide exactly with the data. The strip free energy
lacks logarithmic phase space factors, which are inconsistent with the 
measurement and 
we have included a constant correction term. Each droplet at the 
crystal phase transition tunnels into either one of two possible  
different strip configurations and thus the strip free energy is lowered by 
a term 
$-{\rm ln}(2)$. 

\begin{figure}[tb]
\centerline{ \begin{minipage}[c]{13.2cm}
(a) \psfig{file=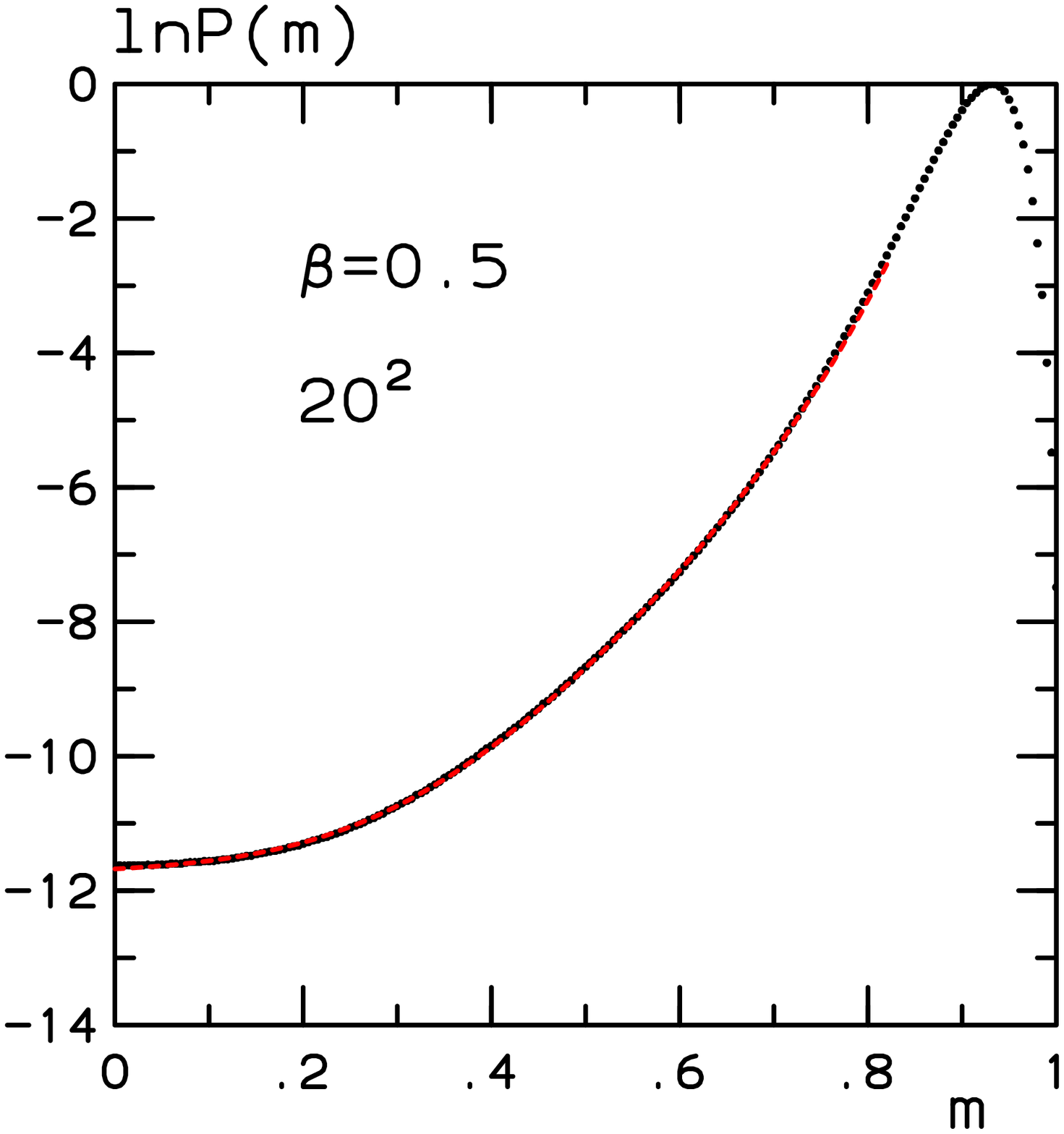,angle=000,width=5.5cm}
(b) \psfig{file=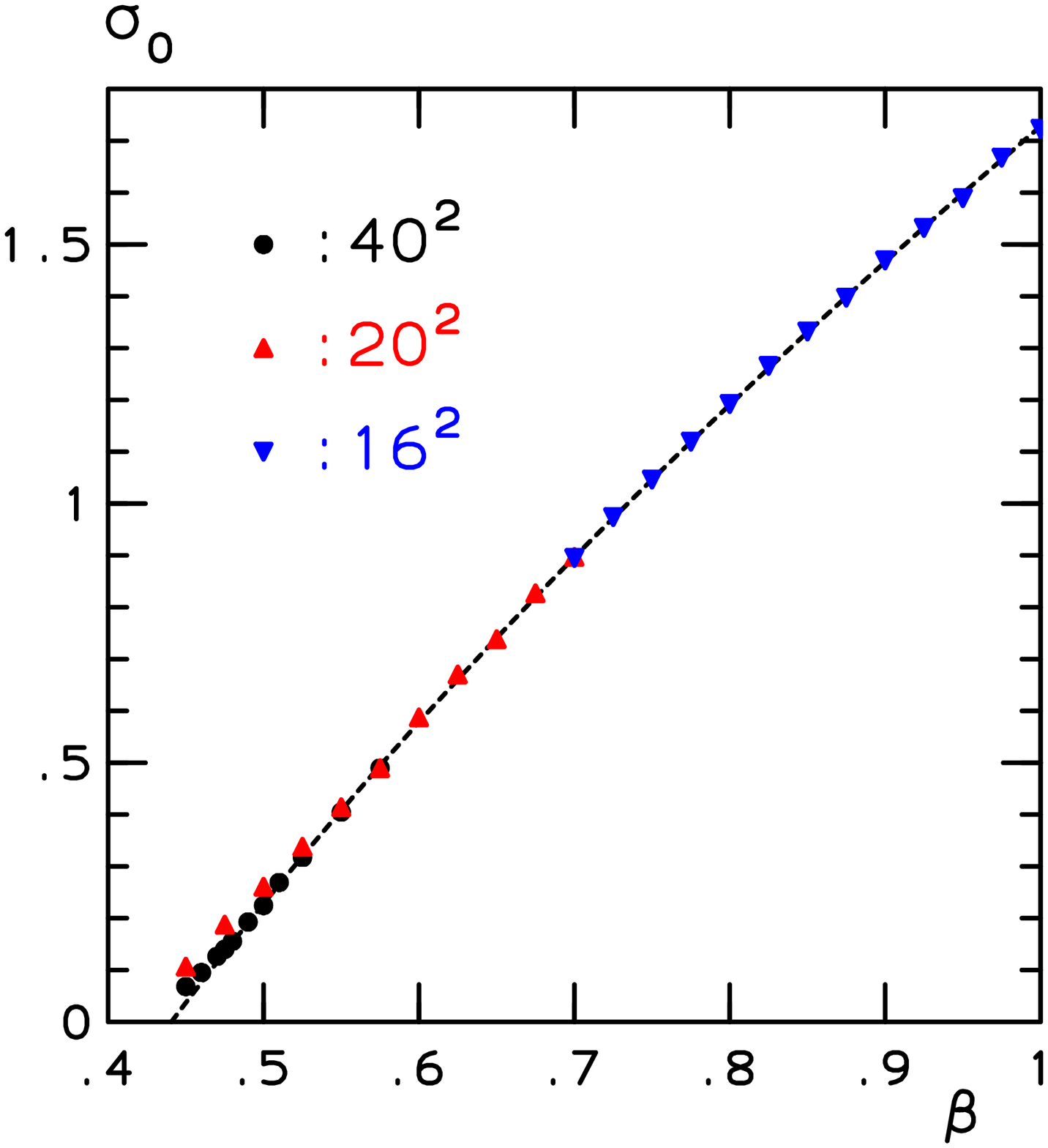,angle=000,width=5.5cm}
    \end{minipage}}
\caption[a]{(a) Logarithmized magnetization density probability distribution 
            data as a function of $m$
            at $\beta=0.5  (T/T_c=0.88)$ on a $20^2$ box and finite size  
rounding fit 
            as explained in the text. Data and the fitted curve lie
            on top of each other. The fitted infinite volume
            interface tension $\sigma_0$ is displayed in (b) as a function  
of $\beta$.
            The curve in (b) corresponds to the exact result.
           }
\label{fig:sigma_from_rounding}
\end{figure}

Droplet and strip free energies of Eqs.~(\ref{drop_new}) and
(\ref{strip_new}) can be used to determine the interface tension
$\sigma_{0}$ from the finite size rounding of the free energy
at the crystal shape transition. If for practical purposes we 
approximate the droplet free energy by
\begin{equation}
F_{droplet} 
\approx \sigma \partial \Omega_0(m) 
+\frac{ \alpha_{GT} }{ \partial \Omega_0(m)^2}
+ {\rm ln}( \frac{\Omega }{ V} )
\end{equation}
we obtain with Eq.~(\ref{d__s_part_func}) 
and with ${\rm ln}Z(m,L)={\rm ln}P_L(m)+const$ at fixed $m_0$ a four parameter
representation of the free energy as a function of the parameters
$\sigma_0,\sigma,\alpha_{GT}$ and $const$, which easily can be fitted. 
A typical set of data for the logarithmized free energy 
at $\beta=0.5$ on a $20^2$ box is displayed in
Fig.~\ref{fig:sigma_from_rounding}a and the fit
for values $m<0.8 m_0$ (droplet and strip phases) lies right on top of the data.
Figure \ref{fig:sigma_from_rounding}b
displays the interface tension $\sigma_0$ as determined from rather small boxes
as a function of $\beta$ in comparison to the exact result, the curve 
in the figure. The agreement is excellent demonstrating, that finite size 
corrections of the free energy are faithfully represented.

\subsection{CONDENSATION PHASE TRANSITION}\label{condeval}

\begin{figure}[t]
\begin{center}
\epsfig{file=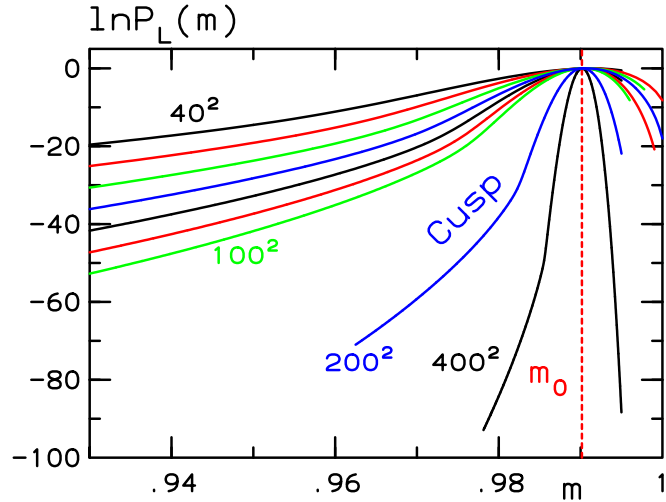,width=8cm,angle=270}
\caption{
Logarithmized magnetization density probability distribution functions $P_L(m)$ 
on toroidal $L^2$ boxes at $\beta=0.7$ for the 2D Ising model. The maxima are 
normalized to values unity in $P_L(m)$. One observes almost Gaussian 
fluctuations in vicinity of $m_0$. Once one approaches the two phase separated 
phase space region, a finite size rounded cusp structure appears, which 
corresponds to the condensation phase transition.}
\label{fig:pofm}
\end{center}
\end{figure}
\begin{figure}[t]
\begin{center}
\epsfig{file=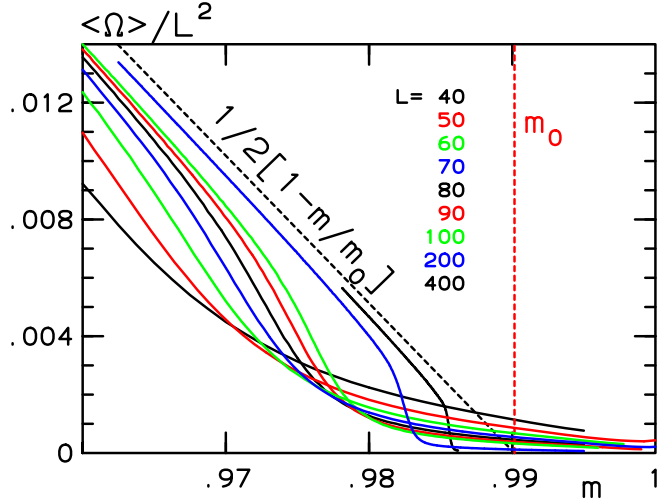,width=8cm,angle=270}
\caption{
Expectation values $<\!\Omega\!>(m)/L^2$ for the minority droplet 
size density in vicinity of the condensation point phase transition for 
$40^2$ up to
$400^2$ boxes. The infinite volume prediction $(1-m/m_0)/2$ corresponds
to the dashed diagonal straight line.} \label{fig:size}
\end{center}
\end{figure}
\begin{figure}[t]
\begin{center}
\epsfig{file=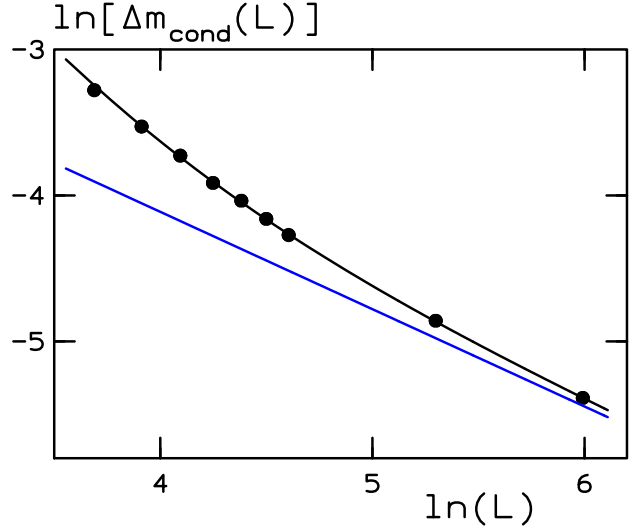,width=8cm,angle=270}
\caption{
Condensation point phase transition transition points at $\beta=0.7$ in the 
2D Ising model. The straight line corresponds to the theoretical prediction,
including Gibbs-Thomson corrections. 
The curve corresponds to a fit to Eq.~(\ref{dmf}) as explained in the text.
}
\label{fig:cond_point_a}
\end{center}
\end{figure}
\begin{figure}[t]
\begin{center}
\epsfig{file=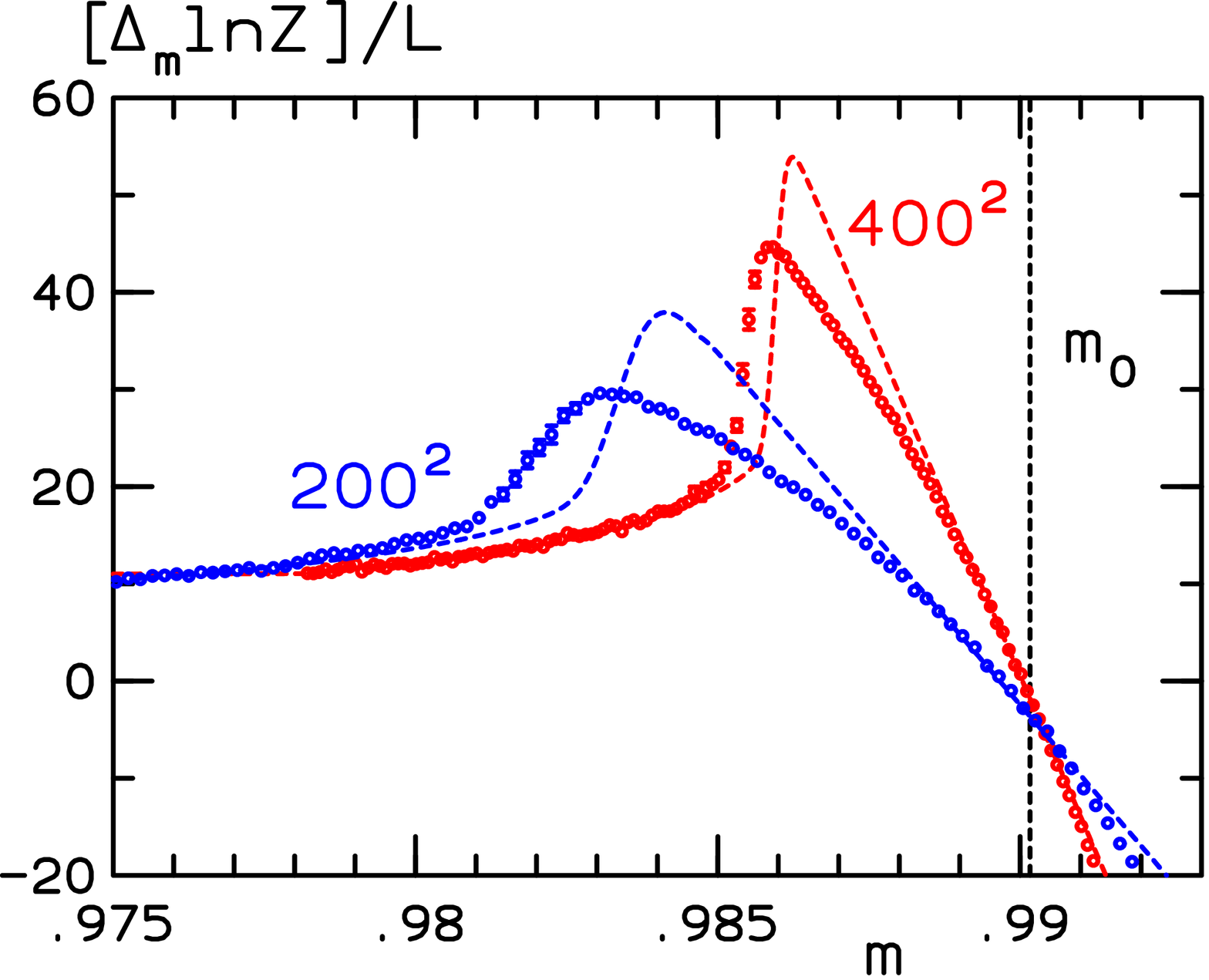,width=8cm,angle=270}
\caption{Finite size rounding of  data for the discrete derivative
$(\Delta_m{\rm ln}Z)/L$ as defined in Eq.~(\ref{deltaz}) for $m$ values in  
vicinity of 
the condensation phase transition on two lattice sizes, $200^2$ 
and $400^2$. The dashed curves correspond to finite size rounding predictions.
Discrepancies are observed for bulk states slightly above the finite system 
phase transitions. The effect is due to the second order Ginzburg-Landau free 
energy expansion.}
\label{fig:z2}
\end{center}
\end{figure}
The phase transition of an extensive minority droplet in the phase separated
phase space region into a gas of small droplets in the bulk phase is studied 
on toroidal lattices. 
In order to provide an overview over the data we display in Fig.~\ref{fig:pofm}
the (logarithmized) probability distribution $P_L(m)$ of the magnetization 
density - the constraint partition function $Z(m,L)$ - as a function of $m$ 
in vicinity of Onsagers magnetization density value $m_0$. 
A finite size rounded cusp structure is visible for values of $m$ slightly 
below $m_0$ and 
corresponds to the position of the condensation phase transition $m_{cond}(L)$. 
The data are obtained on $40^2$ up to $400^2$ boxes with the help of Muca 
simulations, which 
quite similar to simulations with the Wang-Landau algorithm (see section 
\ref{mucamethod}) also 
suffer from exponential down and in fact: it would be quite time consuming 
to obtain data of comparable statistical quality for a $1000^2$ box.

The expectation value for the size of the minority droplet $<\!\Omega\!>(m)$ is 
displayed
in Fig.~\ref{fig:size}. From the peak positions of the $\Omega$ susceptibility
(not displayed in a figure) we determine finite size shifted condensation point 
magnetization
values $m_{cond}(L)$ and calculate the shift due to the finite system 
size $\Delta m_{cond}(L)=m_0-m_{cond}(L)$. The shift is displayed in
Fig.~\ref{fig:cond_point_a} in a double logarithmic scale as a function of $L$. 
Numerical values are contained in table \ref{thetable}.
The condensation point shift of Fig.~\ref{fig:cond_point_a} 
contains a straight line, which corresponds to the theoretical prediction
of classical droplet theory with  Gibbs-Thomson corrections (\ref{shift}).  
While finite size corrections to the theory on small boxes are large, it appears 
perfectly possible, that the data (the solid circles in the figure) approach the 
theoretical prediction on large boxes i.e., for large droplets.
Additional finite size corrections 
are caused by sub-leading free energy contributions
and quite similar to the discussion of section \ref{toreval}, we compare in 
Fig.~\ref{fig:z2} data for the discrete derivative $\Delta_m{\rm ln}Z$
on $200^2$ and $400^2$ boxes with
finite size rounding theory, as predicted by the superposition Ansatz 
(\ref{super_position_condensation}).  Similar Monte-Carlo data for systems of
smaller linear extent were shown in Fig.~10 of \cite{lnr95} for the Ising model
in D=2 and in Figs.~6 and 7 of \cite{fb82} for D=3. The onset of the same  
singularity
as displayed in Fig.~\ref{fig:z2} is clearly visible in those data, despite of the 
presence of strong finite-size rounding.
The Ansatz now uses the precisely known 
droplet free energy of (\ref{drop_new}) and the bulk free energy 
of (\ref{bulkfreeenergy}). The dashed lines of the figure do correspond to 
finite size rounding predictions. It is evident, that the second order  
expansion of 
the  
Ginzburg-Landau free energy is not precise enough. The calculation of 
coefficients $c_n$ 
\be
U_{eff}(m)= \sum_{n=2}^{\infty} c_n (m-m_0)^n
\end{equation}
of an polynomial expansion for the constraint effective potential $U_{eff}(m)$ 
\be
U_{eff}(m)=  {\rm lim}_{L \rightarrow \infty}
[ - \frac{ 1}{ L^2 } {\rm ln} Z(m,L) ]
\end{equation}
in higher powers of $m-m_0$, could provide a better bulk free energy form. 
We will not however pursue this issue here.

\begin{figure}[t]
\begin{center}
\epsfig{file=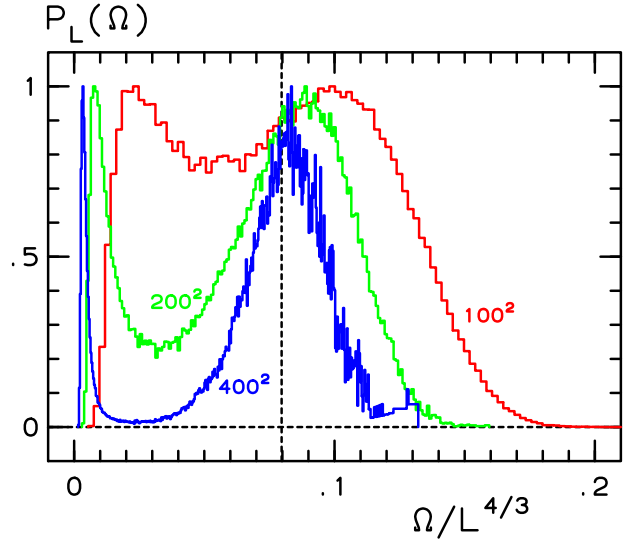,width=8cm,angle=270}
\caption{
Probability distributions $P_L(\Omega)$ of the minority droplet size
at the condensation phase transition. The vertical line corresponds to the
``gap'' prediction of classical droplet theory with Gibbs-Thomson corrections.}
\label{fig:distribution}
\end{center}
\end{figure}

We perform a three parameter fit to $\Delta m_{cond}(L)$ data and parameterize 
additional 
finite size corrections through a single power correction with a free exponent 
value $\beta$ 
\begin{equation}\label{dmf}
\Delta m_{cond}(L)= A_{cond}L^{-{2 / 3}}  + B L^{-\beta} \quad .
\end{equation}
The fit corresponds to the curve in Fig.~\ref{fig:cond_point_a} and at a 
$\chi^2_{dof}$ 
value of $0.25$ we obtain the fit parameters ${B}=23(9)$, $\beta=1.93(11)$ and 
$A_{cond}=0.237(3)$. The fitted value for $A_{cond}$ perfectly agrees with the 
theoretical
prediction $A_{cond}=0.23697...$ (\ref{erg:dgtgl}), which at the condensation 
phase transition
yields support for the validity of classical droplet theory and the presence 
of Gibbs-Thomson 
corrections for large droplets.
\begin{table}[ht]
\centerline{
\begin{tabular}{cccccc}
\hline
 L  & $\Delta m_{cond}$ &  $\Omega_{cxc}/L^{2/3}$ 
& $\Omega_{cxc}$ &  $Q$ & $B_{cxc}/\sigma \sqrt{ 4 \pi \Omega_{cxc}}$\\
\hline
 40 &   0.0376(12)  &       -  &     -    &      -  & - \\
 50 &  0.02936(80)  &       -  &     -    &      -  & - \\
 60 &  0.02405(55)  &       -  &     -    &      -  & - \\
 70 &  0.01995(40)  &       -  &     -    &      -  & - \\
 80 &  0.01766(31)  &       -  &     -    &      -  & - \\
 90 &  0.01559(24)  &       -  &     -    &      -  & -\\
100 &  0.01396(20)  & 0.0976(8) &  45.3(4) & 1.555(25) &  0.013(2)  \\
200 & 0.007762(50)  & 0.0878(5) & 102.7(6) & 1.526(12) &  0.045(4)  \\
400 & 0.004575(12)  & 0.0825(9) & 243.4(6) & 1.518(16) &  0.085(4)  \\
\hline
 $\infty$   & 0.  &  0.07977... & $\infty$  & 1.5  & 0.17404... \\
\hline
\end{tabular}}
\caption[thetable]{Measured observables for the 2D Ising model droplet
condensation phase transition at $\beta=0.7$. The lowest line contains the 
infinite volume
theoretical predictions of classical droplet theory with Gibbs-Thomson 
corrections.}
\label{thetable}
\end{table}
A somewhat more stringent test of the presence of Gibbs-Thomson corrections
can be devised, if at the condensation phase transition the ratio $Q$ of 
(\ref{qvalue}) is calculated. For large droplets one expects, 
that $Q$ equals $Q=3/2$. For magnetization values at - or close to - the 
condensation 
transition, we determine the probability distribution $P_L(\Omega)$ for 
the occurrence of a minority droplet of size $\Omega$ in the magnetization 
constraint partition function at $m$ and tune the magnetization values to the 
point,
where
finite system double peaked probability distributions have equal height. The 
result $P_L(\Omega)$ is displayed in Fig.~\ref{fig:distribution} as a 
function of $\Omega/L^{4/3}$ for boxes $100^2$, $200^2$ and $400^2$. 
\begin{figure}[t]
\begin{center}
\epsfig{file=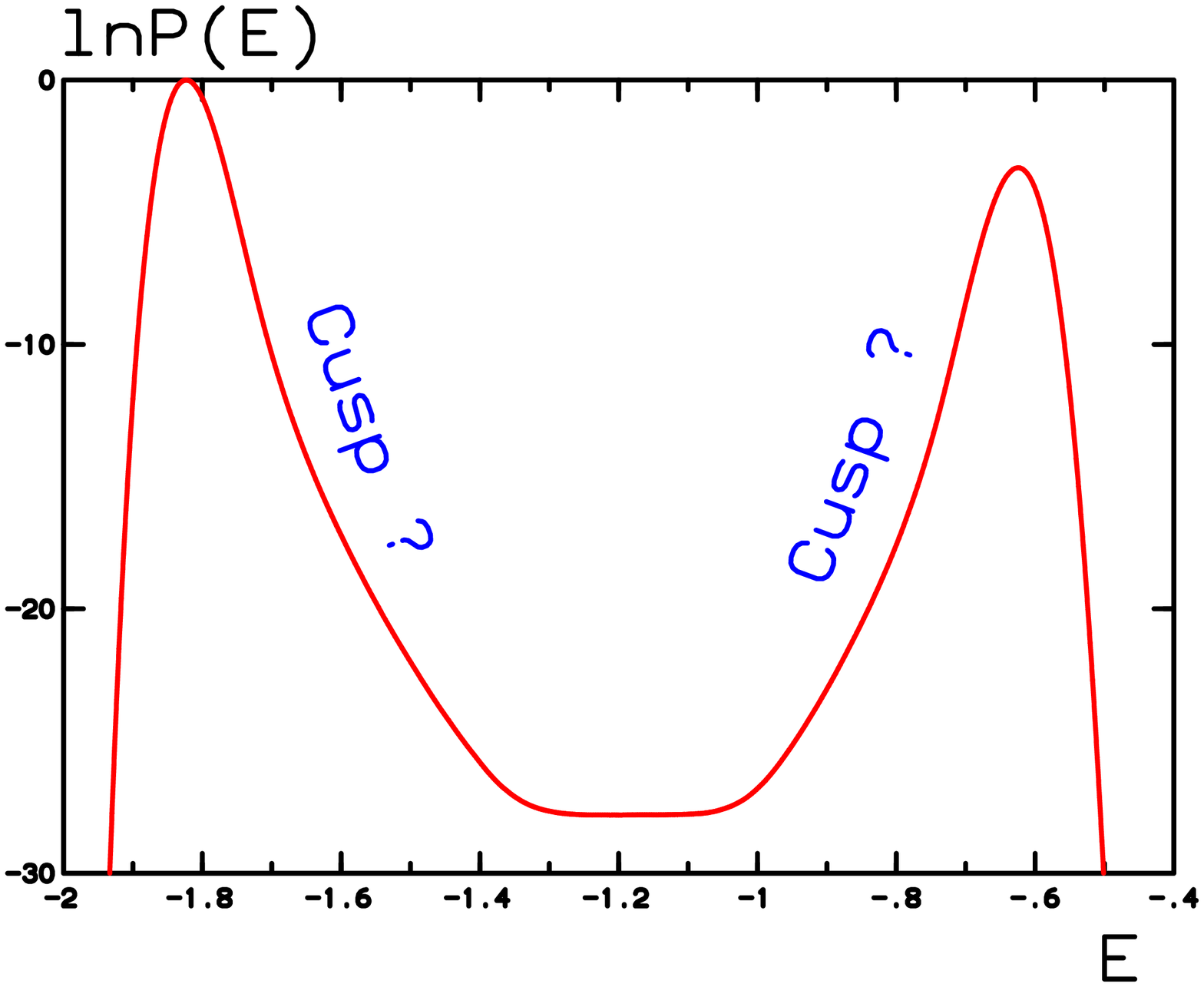,width=8cm,angle=270}
\caption{
Probability distribution $P(E)$ for the energy in the 2D $q=20$ Potts model
on a $70^2$ box. The conjectured positions of condensation phase
transitions are indicated by finite size rounded cusps.}
\label{fig:potts}
\end{center}
\end{figure}
Our numerical simulation demonstrates the existence of double peaks in 
$P_L(\Omega)$. 
There also exists a barrier for states in-between, which increases with  
increasing 
droplet size. These findings provide numerical evidence for the fact, that the 
droplet condensation phase transition of the 2D Ising model is of 
discontinuous nature. Upon fitting Gaussians to the right hand side peaks of 
Fig.~\ref{fig:distribution} we determine 
values for the coexisting droplets size $\Omega_{cxc}$, which are contained
in rows three and four of table \ref{thetable}. 
Values for the ratio $Q$ are given in the fifth row of 
table \ref{thetable}. On the $400^2$ box we find $Q=1.518(16)$, which indeed is 
very close to $Q=3/2$ and thus again provides solid support for the asymptotic 
correctness of 
classical droplet theory and the presence of Gibbs-Thomson corrections.

A very interesting quantity is the nucleation barrier $B_{nucl}$, which at the 
condensation 
phase transition determines the suppression of states in-between the 
equilibrium minority 
droplet  - and the gas of small droplets in the bulk phase. At equal height of 
the probability 
distribution $B_{nucl}$ can be estimated through
\begin{equation}
B_{nucl} \approx \ln \frac{ P_{max} }{ P_{min} } \quad ,
\end{equation}
where values $P_{max}$ and $P_{min}$ denote maxima and minima of $P_L(\Omega)$. 
The last row of table \ref{thetable} contains $B_{nucl}$ values, which are 
given in units of the droplet surface free energy 
$\sigma \sqrt{ 4 \pi \Omega_{cxc}}$.
The data exhibit large finite size effects and most likely one needs much 
larger droplets in order to draw definite conclusions. On the $400^2$ box 
we find a barrier, which is smaller by a factor of $1/2$ than the theoretical 
prediction (\ref{thebarrier}). Chances thus are high, that the   
condensation of large droplets proceeds within Gibbs-Thomson corrected 
classical droplet theory through the formation of intermediate saddle point 
configurations, which are classical in nature also. 

The grand-canonical droplet model of \cite{f67,l67} predicts 
\begin{equation}\label{lang}
\Omega_{crit}= \left[ \frac{(d-1)\hat{\sigma}}{2d m_0  h} \right]^d \quad\quad d\geq  
2\quad.
\end{equation}
for the volume $\Omega_{crit}$ of the 
instable critical droplet as a function of the ordering field $h$.
Taking into account that the 
the surface tension $\hat{\sigma}=\sqrt{4\pi}\sigma$ of \cite{f67,l67}  
differs from ours by a factor $\sqrt{4\pi}$ one obtains
\begin{equation}\label{critical_droplet}
\Omega_{crit}=\frac{\pi\sigma^2}{4 m_0^2  h^2} \quad
\end{equation}
in $d=2$. Using Eq. (\ref{dropsize}) and $\Delta m=h/2c_2$ we 
compare $\Omega_{crit}$ with the volume
of the equilibrium droplet at the condensation phase transition
\begin{equation}
\Omega_{cxc}= \frac{A_{cond}^3}{3 m_0\Delta m_{cond}^2} \
=9\frac{\pi \sigma^2}{4 m_0^2h_{cond}^2} \quad ,
\end{equation}
which has the same field dependence as (\ref{critical_droplet}). 
$\Omega_{cxc}$ is a 
factor nine times larger than $\Omega_{crit}$. 
As one can see in 
Fig.~\ref{fig:distribution}
this is in 
accordance with our simulations. The droplet volume ratios 
of the stable droplet at coexistence over the critical droplet 
at the saddle point approximately have values $2,3$ and $4$ for the considered 
lattice sizes. It is very well possible, that these volume ratios approach 
the predicted value $9$ in the infinite volume limit.

Finally we want to discuss how the free energy $F_L=\ln P_L(M)$
in general and the observed transitions in particular affect the dynamics of the 
decay of a metastable state. As it stands $F(L)$ describes the equilibrium state 
for conserved order parameter $m$. It tells us  for example whether adatoms with
a given density $\Delta m$ on a surface of size $L^2$ form a large single 
crystal island or a homogenous adatom gas. For systems with conserved order 
parameter, like binary alloys the typical experimental setting is a rapid quench
from a uniform state at high temperature into the two phase region, followed by 
nucleation of droplets or spinodal decomposition.
In magnetic switching experiments \cite{rklrn97} a  
system with a non-conserved order parameter is driven from one 
equilibrium bulk state to the other by  
an external magnetic field $h$ and pending on the field strength 
one observes two different tunneling regimes:
In the stochastic regime at small fields tunneling proceeds via the 
nucleation of a single droplet and as the process is rare 
one observes large fluctuations in tunneling times, if the 
experiment is repeated
many times. At large fields 
multi-droplet states dominate the intermediate configurations 
of the early tunneling process.
The multi-droplet nature of the tunneling in this case reduces fluctuations
of tunneling times. The crossover region in-between both regimes was called the 
dynamic spinodal \cite{rtms94}. 

The decay of supersaturated bulk states in any case proceeds  
via the formation of a single droplet - or via an ensemble of several critical 
droplets. Let us consider a subsystem of supersaturated matter 
of linear extent $\xi$ and volume $\xi^2$ 
at fixed finite $h$, which is embedded into a 
larger system of finite linear size $L$ and volume $L^2$. 
The properties of critical droplets in 
vicinity of the condensation phase transition 
determine certain aspects of the tunneling processes, which
can be activated within the subsystem. Most notably and due to 
the existence of the condensation phase transition we observe, that volumes of critical 
droplets $\Omega_{crit}$ cannot be 
arbitrary small for given subsystem size $\xi$. 
Their values are bounded from below through a $\xi$ dependent 
bound, which is saturated 
at the condensation phase transition 
\begin{equation}\label{inequality}
\Omega_{crit} \ge \frac{1}{27 m_0A_{cond}} ~ \xi^{4/3} \quad
\end{equation}
with $A_{cond}$ as given in Eq. (\ref{erg:dgtgl}). 
The volume of critical droplets according to
Eq. (\ref{critical_droplet}) is proportional to $h^{-2}$
and can have any value. At small values of $h$ the critical droplet volume
is large and the inequality of Eq. (\ref{inequality}) is satisfied even 
for values $\xi=L$. In this region stochastic tunneling 
takes place with the nucleation of a single droplet. 
On the other hand, at large values of $h$ the critical droplet volume
is small and the inequality of Eq. (\ref{inequality}) only is satisfied 
if $\xi<L$. Consequently tunneling proceeds within
many subsystems of size $\xi$ i.e., multi-droplet
states are  encountered. For finite size systems with linear extent $L$ and
volume $L^2$ we determine the magnetic field $h_{DS}$, where the crossover from
single- to multi-droplet decay takes place, as
\begin{equation}\label{dynamic_spinodal}
h_{DS} = 2 c_2 A_{cond}~ L^{-2/3} \quad
\end{equation}
and for values $h>h_{DS}$ we find the subsystem linear size 
\begin{equation}
\xi(h)=[\frac{{2 c_2 A_{cond}}}{h}]^{3/2} 
\end{equation}
with $\xi(h=h_{DS})=L$. The interplay of the condensation
phase transition with the dynamic spinodal \cite{rtms94} was already
noted in \cite{lnr95}. These authors argued however in favor 
of a logarithmic $L$ - dependency $(\ln L/L)^2$ instead of the $L^{-2/3}$  
dependence of the critical field $h_{DS}$. Recent rigorous arguments \cite{biv00,bck02}
on the location of the condensation phase transition  exclude such leading 
logarithmic corrections in the finite size behavior of $F_L(m)$. Our simulation reveals sizable sub-leading finite size 
corrections for the considered lattice sizes, which deserve further studies.  
With the assumption, that the multi-droplet  
physics necessary to derive the location of the dynamic spinodal is already contained   
in the free energy landscape of the restricted magnetization equilibrium partition
function, logarithmic terms should be absent in the leading $L$ dependence of $h_{DS}$.  Further efforts are necessary to check, whether this assumption holds true or wether the
dynamical effects invoked in \cite{rtms94} change the $L$ dependence of $h_{DS}$.

\section{Conclusion and Outlook}

The constraint magnetization partition function of the  2D
Ising model, which we studied at inverse temperature $\beta=0.7$, 
hosts a variety of thermodynamic 
singularities, which qualify as genuine first-order phase transitions
with free energy barriers, that diverge in the thermodynamic limit. The phase
space on the torus splits into five disconnected sectors: bulk, droplet
and strip phases ($+/-$ symmetric states counted separately) and, on  
cube-surface  
lattices the number of phases is even higher, nine in total.
On toroidal and cube-surface lattices we observe shape transitions,
which strongly depend on the choice of the lattice manifold and, which 
for finite systems only can be avoided, if the model could be formulated on a 
perfect sphere. Since no such regularization is known, we are faced with
discontinuous behavior for certain quantities and with free energy barriers, 
which are 
proportional to the system size $L$. The barrier value and the position of the 
crystal shape transition 
on the torus (from a droplet to a strip) and the positions 
and barriers of the ``corner occupying'' droplet shape transitions on the 
cube-surface, 
all are very well described by a classical droplet description. 

The algorithmic performance of multicanonical ensemble simulations suffers 
from the existence 
of such shape transitions. Barriers at the discontinuous phase transitions can 
not be removed and result into exponential slowing down. 
The earlier general conjecture of random walk behavior for multicanonical 
ensemble 
simulations in applications to first-order phase transitions is falsified for 
a special case. 
This interesting algorithmic fact untill now has escaped detection, just because 
free energy barriers in most applications of multicanonical ensemble 
simulations - though present - were not particularly large. There is no 
doubt, that energy driven 
first-order phase transitions on toroidal lattices with periodic boundaries 
possess similar crystal shape transitions e.g., from a droplet to a strip. 
In higher dimensions 
than $d=2$, e.g. in $d=3$, one faces the additional fact, that two phase 
separation on toroidal boxes in intermediate stages also proceeds with the 
formation of cylinders, which adds additional shape transitions to the 
scenario. We expect, that the 
performance degradation, which is observed in multicanonical ensemble 
simulations, is a 
general property of broad histogram sampling methods for phase separated 
systems 
and, that Wang-Landau density of states updating also is affected.

For toroidal lattices we have obtained a precise finite size
parameterization for droplet and strip free energies - and for the finite 
size energy rounding at the crystal shape transition. Our finding predicts 
the existence of logarithmic phase space factors in the droplet volume, which
in addition to Gibbs-Thomson and capillary wave fluctuation corrections, 
contribute to the droplet free energy. Such terms reflect the fact, 
that droplets at fixed  magnetization may fluctuate to any spatial position 
and, for the considered droplet 
sizes at the crystal shape transition these corrections are actually larger, 
than e.g. 
Gibbs-Thomson corrections. A Tolman correction to the classical droplet free 
energy could not be observed, confirming the prediction of \cite{fw84} that 
the amplitude of this correction vanished for systems like the Ising model, 
which are symmetric under the interchange of the two phases. 
We have conjectured, that translational invariance for strip states is broken, 
because similar logarithmic phase space factors are absent for the strip free 
energy. It would be quite 
interesting to study the dependence of this effect on temperature and dimension.

The phase separated region of the 2D Ising model is bounded by a condensation 
phase transition, which as we checked at $\beta=0.7$
is of discontinuous nature. Within the scope of the present paper we 
worked out the consequences of a simple theoretical model in the one
droplet sector, which is based on classical droplet theory and 
Gibbs-Thomson corrections. Gibbs-Thomson corrected classical droplet theory 
decomposes the free energy into the classical contribution of the 
droplet, adding fluctuations through the expansion of the Ginzburg-Landau 
free energy, which in this paper was considered up to second order. 
This theory performs surprisingly well for large droplets. The finite size 
condensation phase transition shift, the gap in the minority droplet size, 
the ratio of the shift over the gap and the nucleation barrier - all seem to 
approach the second order 
Ginzburg-Landau free energy theoretical prediction. Future studies should 
answer the important theoretical question, whether all of the observed 
corrections for large, small and 
smallest droplets can be incorporated through a higher order expansion of the 
Ginzburg-Landau free energy, which as already mentioned requires
precise knowledge on the shape of the constraint effective potential in 
vicinity of the bulk. We have also presented an argument relating
the condensation phase transition to the location of the dynamic spinodal, which
is of relevance for magnetic field switching experiments.

Thermodynamic singularities associated with the condensation of droplets in 
models with phase separation are expected to exist in any dimension - and 
also for temperature driven first-order phase transitions. To our knowledge 
such transitions have not yet been studied
with similar methods in numerical simulations and - for most of the time - the 
effect has plainly been overlooked. For purposes of illustration we display in 
Fig.~\ref{fig:potts} 
the energy probability distribution function $P(E)$ on a $70^2$ box 
from \cite{potts_distri} in the
two-dimensional $q=20$ 
Potts model at the transition temperature. The 
conjectured positions of two asymmetric condensation phase 
transitions again is indicated by finite size 
rounded cusps. The thermodynamic properties of these transitions, like their 
order and 
their nature in terms of droplet and fluctuation degrees have not yet been  
studied, 
neither in dimension two, nor in higher dimensions.

\section*{Acknowledgments}

We like to thank W. Selke for his remarks, that lead to the study of the 
condensation phase transition and P.A. Rikvold for a critical reading of the manuscript and
many suggestions to improve it. J.H. wants to thank M.E. Fisher and R.K.P. Zia 
for useful discussions. T.N. acknowledges an earlier useful discussion with
K. Binder.

\section*{Note added in proof}
After completion of this work we became aware of recent work on details of the condensation transition by K. Binder  (Physica A {\bf 319}:99 (2003), cond-mat/0303651), K. Binder and  coworkers (cond-mat/0303642) and by M. Biskup, L. Chayes and R. Koteck\'{y} (cond-mat/0302373, math.PR/0212300, math-ph/0302031).


\end{document}